\documentclass[preliminary,copyright,creativecommons,sharealike,noncommercial]{eptcs}
\providecommand{\event}{DEVCON V}
\pdfoutput=1

\usepackage[toc,page]{appendix}
\usepackage{mathtools} 
\usepackage{amssymb} 
\usepackage{amsthm} 
\usepackage{stmaryrd} 

\usepackage{relsize} 
\usepackage{microtype} 
\usepackage{multicol} 
\usepackage{csquotes} 
\usepackage{xspace}
\usepackage{hyperref} 
\usepackage[nocompress]{cite} 
\usepackage[hyperpageref]{backref} 
\renewcommand*{\backref}[1]{}
\renewcommand*{\backrefalt}[4]{%
	\ifcase #1 (Not cited.)%
	\or        (Cited on page~#2.)%
	\else      (Cited on pages~#2.)%
	\fi}

\usepackage{graphicx} 
\usepackage{subcaption} 
\usepackage{wrapfig} 
\usepackage[usenames,dvipsnames]{xcolor} 
\usepackage{tikz} 
\usetikzlibrary{
	math,
	decorations.markings,
	positioning,
  arrows,
  intersections,
  shapes,
  shapes.misc,
	automata,
	petri,
	decorations,
	backgrounds,
	calc,
	fit,
	quotes
  }

  \usepackage{circuitikz}

  \newcounter{theoremUnified} 
  \def\thetheoremUnified{\arabic{section}} 
  \numberwithin{theoremUnified}{section} 
  \numberwithin{theoremUnified}{section} 

  \newtheoremstyle{plainStyle} 
  {2mm} 
  {2mm} 
  {} 
  {} 
  {\bfseries} 
  {.} 
  {.5em} 
  {} 

  \newtheoremstyle{italicStyle} 
  {2mm} 
  {2mm} 
  {\itshape} 
  {} 
  {\bfseries} 
  {.} 
  {.5em} 
  {} 

  \theoremstyle{plainStyle} 
    
      \AfterEndEnvironment{example}{\noindent\ignorespaces}
    
      \AfterEndEnvironment{fact}{\noindent\ignorespaces}
    
      \AfterEndEnvironment{examplehard}{\noindent\ignorespaces}
    
      \AfterEndEnvironment{remark}{\noindent\ignorespaces}
    
    \AfterEndEnvironment{remarkhard}{\noindent\ignorespaces}
  \theoremstyle{italicStyle} 
    \newtheorem{definition}[theoremUnified]{Definition}
      \AfterEndEnvironment{definition}{\noindent\ignorespaces}
    
      \AfterEndEnvironment{proposition}{\noindent\ignorespaces}
    \newtheorem{lemma}[theoremUnified]{Lemma}
      \AfterEndEnvironment{lemma}{\noindent\ignorespaces}
    \newtheorem{theorem}[theoremUnified]{Theorem}
      \AfterEndEnvironment{theorem}{\noindent\ignorespaces}
    
      \AfterEndEnvironment{corollary}{\noindent\ignorespaces}

\newcommand{\Naturals}{\mathbb{N}} 
\newcommand{\Bool}{\mathbb{B}} 

\newcommand{\Obj}[1]{\operatorname{Obj} \, #1} 
\newcommand{\Mor}[1]{\operatorname{Mor} \, #1} 

\newcommand{\Snark}[1]{\operatorname{Sn}(#1)} 

\newcommand{\Source}[2]{\operatorname{s}_{#1}(#2)} 
\newcommand{\Target}[2]{\operatorname{t}_{#1}(#2)} 
\newcommand{\Id}[1]{id_{#1}} 

\newcommand{\Free}[1]{\mathfrak{F}(#1)}
\newcommand{\UnFree}[1]{\mathfrak{U}(#1)}

\newcommand{\Tensor}{\otimes} 

\newcommand{\Bfun}{\Bool_\textbf{fun}} 
\newcommand{\Bcirc}{\Bool_\textbf{circ}} 
\newcommand{\Bkp}{\Bool_\textbf{KP}} 
\newcommand{\Bzkp}[1]{\Bool^{#1}_\textbf{ZKP}} 

\newcommand{\Bpath}[1]{\Bool^{#1}_{\textbf{path}}} 
\newcommand{\Count}{\textbf{Count}} 
\newcommand{\Bgraph}[1]{\Bool^{#1}_{\Graph}} 

\newcommand{\Graph}{\textbf{Graph}} 
\newcommand{\Cat}{\textbf{Cat}} 

\newcommand{\NAND}{\ensuremath{\texttt{NAND}}\xspace}
\newcommand{\AND}{\ensuremath{\texttt{AND}}\xspace}
\newcommand{\NOT}{\ensuremath{\texttt{NOT}}\xspace}
\newcommand{\OR}{\ensuremath{\texttt{OR}}\xspace}

\newcommand{\COPY}{\ensuremath{\texttt{COPY}}\xspace}
\newcommand{\TRUE}{\ensuremath{\texttt{TRUE}}\xspace}
\newcommand{\FALSE}{\ensuremath{\texttt{FALSE}}\xspace}
\newcommand{\Zero}{\ensuremath{\textbf{0}}\xspace}

\newcommand{\ANDSym}{
  \scalebox{0.3}{
    \tikz[baseline=-10pt] \node[thick, american and port] (char) {};
  }
}

\newcommand{\ORSym}{
  \scalebox{0.3}{
    \tikz[baseline=-10pt] \node[thick, american or port] (char) {};
  }
}
\newcommand{\COPYSym}{
  \scalebox{0.3}{
    \tikz[baseline=-8pt]{
      \node[thick, draw, circle, radius=2pt] (copy) at (0,0) {};
      \node (in) at (-1,0) {};
      \node (out1) at (1,0.5) {};            
      \node (out2) at (1,-0.5) {};
      \draw[thick, -] (in.center) to (copy);
      \draw[thick, -, bend left] (copy) to (out1.center);
      \draw[thick, -, bend right] (copy) to (out2.center);
    } 
  }
}

\newcommand{\MATCHSym}{
  \scalebox{0.3}{
    \tikz[baseline=-8pt]{
      \node[thick, draw, fill=gray, circle, radius=2pt] (copy) at (0,0) {};
      \node (in) at (1,0) {};
      \node (out1) at (-1,0.5) {};            
      \node (out2) at (-1,-0.5) {};
      \draw[thick, -, dotted] (in.center) to (copy);
      \draw[thick, -, bend right] (copy) to (out1.center);
      \draw[thick, -, bend left] (copy) to (out2.center);
    } 
  }
}

\title{Mapping Finite State Machines to zk-SNARKs \\ Using Category Theory}
\author{
	Fabrizio Genovese \hspace{1em} Andre Knispel \hspace{1em} Joshua Fitzgerald \\
	Statebox Team\\
	\texttt{research@statebox.io}
}

\def\titlerunning{Mapping finite state machines to zk-SNARKs}
\def\authorrunning{F. Genovese, A.Knispel \& J. Fitzgerald}

\begin{document}
\maketitle
\begin{abstract}
  We provide a categorical procedure to turn graphs corresponding to 
  state spaces of finite state machines into boolean circuits, leveraging on the fact that 
  boolean circuits can be easily turned into zk-SNARKs. Our 
  circuits verify that a given sequence of edges and nodes is indeed a 
  path in the graph they represent. We then generalize to circuits verifying 
  paths in arbitrary graphs. We prove that all of our correspondences are 
  pseudofunctorial, and behave nicely with respect to each other.
\end{abstract}
\section{Introduction}\label{sec:introduction}
Lately, especially due to the advent of smart contracts 
in business applications, there has been a renewed interest towards 
classical results in theoretical computer science.

Smart contracts, especially if hosted on the blockchain~\cite{Nakamoto2008, Buterin2014}, are immutable
pieces of code, more often than not used to manage money. These are 
very good reasons to look into ways of writing smart contracts  
that are reliable, easy to analyze and correct-by-construction. These requirements
 renewed interest in formal models of computation~\cite{Rosu2018}: In particular, 
 \emph{finite state machines (FSMs)}~\cite{Mavridou2017} are considered easy 
 to implement, well structured, and make possible to prove 
 properties of the computations being performed. 

On the other hand, blockchain also spawned a renewed interest for 
cyptography, both for security, privacy, and space reasons: It is 
paramount for blockchains to be cryptographically secure, if they are 
meant to work as exchanges of valuables of any sort (such as digital currency).
As for privacy, cryptographic tools such as \emph{zero-knowledge 
succinct non-interactive argument of knowledge (zk-SNARKs)}~\cite{Ben-Sasson2013} allow for 
the verification that an information is correct without actually revealing nothing 
about the information itself. This has been used, for instance, by ZCash~\cite{Hopwood2019} to 
implement private blockchain transactions: Here, transacting parties submit 
zk-SNARKs of their transactions to the blockchain, and miners then verify that a given 
transaction followed the rules of the protocol by verifying the zk-SNARK, 
without gaining any information about who-sent-money-to-who, and 
how much. 
Regarding space, it has to be noted that by design blockchains tend to 
grow indefinitely space-wise as new blocks keep being added to the chain~\cite{MacManus2018}. 
This is a serious issue, since new nodes are either forced to download many 
gigabytes of data to sync with the network or they have to require the current state 
of the chain from another node (which requires trusting the node) and then start 
syncing from there. This ``trust Vs. feasibility'' issue can be resolved by \emph{recursive 
zk-SNARKs}~\cite{Ben-Sasson2017}, which can be used to verify, using just a few Kilobytes, that the current 
state received by a node is valid. Such applications look very promising especially in 
contexts such as blockchain applied to the Internet of Things~\cite{Leiba2018}.

In this work, we put together formal models of computation and 
cryptography, providing a categorical way to turn finite state machines into zk-SNARKs 
that verify how a sequence of inputs leading to a state change follows the 
rules specified by the finite state machine itself. To do this, we bypass the problem 
of modelling cryptographical primitives categorically, using the 
fact that boolean circuits can be easily turned into zk-SNARKs by already 
available techniques. 

We proceed as follows: In Section~\ref{sec: the categories bfun, bcirc, bkp} we 
define boolean circuits from a categorical perspective. 
In Section~\ref{sec: finite state machines} we briefly 
explain the links between finite state machines and free categories. 
In Section~\ref{sec: turning executions into circuits} 
we show how to turn a given sequence of state changes 
for a given finite state machine into a boolean circuit. We then obtain a 
boolean circuit which verifies arbitrary sequences up to 
a given length, and show how it can be turned into a zk-SNARK. 
In Section~\ref{sec: abstracting over graphs} we generalize to 
circuits which accept the specification defining a finite state machine 
as input, thus attaining full privacy. In Section~\ref{sec: conclusion} 
we conclude by defining directions of future work.

\section[The categories Bfun, Bcirc, Bkp]{The categories $\Bfun$, $\Bcirc$, $\Bkp$}
\label{sec: the categories bfun, bcirc, bkp}
\begin{definition}
  A \emph{boolean function} is a function $\Bool^n \to \Bool^m$, for naturals $m,
  n$. We denote with $\Bfun$ \emph{the category of boolean functions}, having
  $\Bool^n$, for each natural $n$ as objects, and boolean functions as
  morphisms. Composition is the usual function composition. This category is
  clearly symmetric monoidal, with $\Bool^0$ as unit, and the usual product of
  functions as product.
\end{definition}
We want to give a categorical description of boolean circuits, which are wirings
of logical gates that compute a boolean function. The way these circuits are
wired is classically modeled by directed acyclic graphs, however we can model
them as morphisms in a monoidal category. First, we need to choose a set of
gates:
\begin{definition}
  A \emph{set of gates} consists of a family of sets $G_{n,m}$, and a family of
  functions $\texttt{int}_{n,m} : G_{n,m} \to (\Bool^n \to \Bool^m)$.
\end{definition}
\begin{definition}
  Let $G$ be a set of gates. We denote with $\Bcirc^G$ \emph{the category of
    boolean circuits with gates in $G$}, which is the free symmetric strict
  monoidal category generated by one object, denoted $X$, and morphisms $m_g :
  X^n \to X^m$ corresponding to elements $g$ of $G_{n,m}$. We will often use $X^n$
  to denote the $n$-fold monoidal product of $X$, and $X^0$ to denote the
  monoidal unit. For more information about how to generate a free symmetric
  strict monoidal category from a set of object and morphism generators,
  see~\cite{Genovese2019}.
\end{definition}
From there, we get a functor that maps a boolean circuit to the function it computes:
\begin{lemma}\label{lem: monoidal functor Bcirc Bfun}
  There exists a strict monoidal functor $\texttt{ext}^G: \Bcirc^G \to \Bfun$ sending
  the generating morphisms $m_g$ to the function $\texttt{int}(g)$.
\end{lemma}
For our purposes, it is necessary that every boolean function can be computed by
a boolean circuit, i.e. that the functor $\texttt{ext}^G$ is full.
\begin{definition}
  A set of gates is called \emph{functionally complete} if $\texttt{ext}^G$ is full.
\end{definition}
This is a reformulation of the classical definition of functional completeness
(see~\cite{Henderton2001}). An important distinction between our formalism an
the classic one is that we have to explicitly add the constant gates and a \COPY
gate.
\begin{lemma}
  The set of circuits consisting of \NAND, \COPY, \TRUE and \FALSE is
  functionally complete. We denote the morphisms they generate as follows:
  
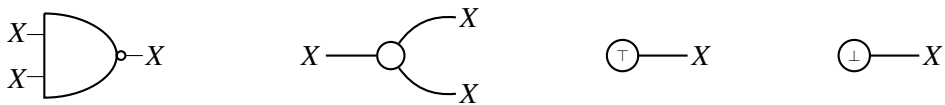
\begin{equation*}
    \tikz[baseline]{
      \node[american nand port] (nand) at (1.5,0) {};
      \node at ([xshift=5pt]nand.out) {$X$};
      \node at (0,0.3) {$X$} ;
      \node at (0,-0.3) {$X$} ;
    }
    \qquad\qquad
    \tikz[baseline]{
      \node[thick, draw, circle, radius=2pt] (copy) at (0,0) {};
      \node[label={[label distance=-10pt]180:$X$}] (in) at (-1,0) {};
      \node[label={[label distance=-10pt]0:$X$}]  (out1) at (1,0.5) {};            
      \node[label={[label distance=-10pt]0:$X$}]   (out2) at (1,-0.5) {};
      \draw[thick, -] (in) to (copy);
      \draw[thick, -, bend left] (copy) to (out1);
      \draw[thick, -, bend right] (copy) to (out2);
    }
    \qquad\qquad
    \tikz[baseline]{
      \node[thick, draw, circle, radius=2pt, inner sep = 2pt] (true) at (0,0) {\tiny{$\top$}};
      \node[label={[label distance=-10pt]0:$X$}] (out) at (1,0) {};        
      \draw[thick, -] (copy) to (out);
    }
    \qquad\qquad
    \tikz[baseline]{
      \node[thick, draw, circle, radius=2pt, inner sep = 2pt] (true) at (0,0) {\tiny{$\bot$}};
      \node[label={[label distance=-10pt]0:$X$}] (out) at (1,0) {};            
      \draw[thick, -] (copy) to (out);
    }
  \end{equation*}
\end{lemma}
For the remainder of this paper we fix a functionally complete set of circuits
and omit the index referring to it. We will refer to specific gates, such as
$\ORSym$ (\emph{\OR}) or $\ANDSym$ (\emph{\AND}): In our setting, these are just
syntactic sugar for the opportune circuit that simulates them.

We also need a category that models boolean circuits that allow possibly
incorrect inputs as well as extra inputs that are aggregated when morphisms are
composed:
\begin{definition}\label{def: definition of bkp}
  We denote with $\Bkp$ the \emph{bicategory of knowledge proof circuits} defined 
  as follows:
  \begin{itemize}
    \item $\Obj{\Bkp} := \Obj{\Bcirc}$;
    \item $\Mor{\Bkp}(A,B) := \Mor{\Bcirc}(A \Tensor X^n, X \Tensor B )$, for all $n \in \Naturals$.
    We depict morphisms as shown below; the $X^n$ and $X$ wires are ``silent'' with respect to 
    our categorical structure, so we depict them dashed and dotted, respectively:
    \begin{equation*}
    \begin{tikzpicture}
      \node[draw, minimum height=30, minimum width=5, rounded corners=2, thick] (f) at (-1,0.5) {$f$};
      \node (A) at (-2,0.75) {$A$};
      \node (Xn0) at (-2,0.25) {$X^{n}$};
      \node (Bf) at (0,0.75) {$X$};
      \node (B) at (0,0.25) {$B$};

      \draw[thick] (A) -- (-1.25,0.75);
      \draw[dashed,thick] (Xn0) -- (-1.25,0.25);

      \draw[dotted,thick] (-0.75,0.75) -- (Bf);
      \draw[thick] (-0.75,0.25) -- (B);
    \end{tikzpicture}  
    \end{equation*}
    \item $\Id{A} := \top \Tensor \Id{A}: A \Tensor X^0 \to X \Tensor A$. Identites are
    depicted as follows:
    \begin{equation*}
      \begin{tikzpicture}
        \node[draw, dashed, minimum height=30, minimum width=20, rounded corners=2, thick] (f) at (-1,0.5) {};
        \node[thick, draw, circle, radius=2pt, inner sep = 2pt] (true) at (-1,0.75) {\tiny{$\top$}};
        \node (A) at (-2,0.25) {$A$};
        \node (Bf) at (0,0.75) {$X$};
        \node (B) at (0,0.25) {$A$};
  
        \draw[thick] (A) -- (B);
  
        \draw[dotted, thick] (-0.75,0.75) -- (Bf);
      \end{tikzpicture}  
      \end{equation*}
    \item $\Mor{\Bkp}(A,B)(f,g) = \begin{cases}
      \{*\} & \text{ iff } \texttt{ext}f = \texttt{ext}g;\\
      \,\,\,\, \emptyset & \text{ otherwise}
    \end{cases}$
    \item Given $f : A \to B$ and $g: B \to C$, corresponding to 
    morphisms of $\Bcirc$ $A \Tensor X^{n_0} \to X \Tensor B$ and 
    $B \Tensor X^{n_1} \to X \Tensor C$, respectively, 
    we set $f;g$ to be the morphism 
    $(f \Tensor \Id{X^{n_1}});(\Id{X} \Tensor g);(\ANDSym \Tensor \Id{C})$.
    Composition is depicted graphically as follows:
    
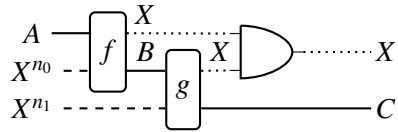
\begin{equation*}
      \begin{tikzpicture}
        \ctikzset{tripoles/american and port/input height=0.7};
        \ctikzset{tripoles/american and port/height=.5};
        \ctikzset{tripoles/american and port/width=.7};
        \node[draw, minimum height=30, minimum width=5, rounded corners=2, thick] (f) at (-1,0.5) {$f$};
        \node[draw, minimum height=30, minimum width=5, rounded corners=2, thick] (g) at (0,0) {$g$};
        \node[american and port] (and) at (1.5,0.5) {};

        \node (A) at (-2,0.75) {$A$};
        \node (Xn0) at (-2,0.25) {$X^{n_0}$};
        \node (Xn1) at (-2,-0.25) {$X^{n_1}$};

        \node (Bf) at (-0.5,1) {$X$};
        \node (B) at (-0.5,0.5) {$B$};

        \node (Bg) at (0.5,0.5) {$X$};
        
        \node (X) at (2.7,0.5) {$X$};
        \node (C) at (2.7,-0.25) {$C$};

        \draw[thick] (A) -- (-1.25,0.75);
        \draw[dashed, thick] (Xn0) -- (-1.25,0.25);
        \draw[dashed, thick] (Xn1) -- (-0.24,-0.25);

        \draw[dotted, thick] (-0.75,0.75) -- (and.in 1);
        \draw[thick] (-0.75,0.25) -- (-0.24,0.25);

        \draw[dotted, thick] (0.24,0.25) -- (and.in 2);

        \draw[dotted, thick] (and.out) -- (2.5,0.5);
        \draw[thick] (0.24,-0.25) -- (2.5,-0.25);
      \end{tikzpicture}
    \end{equation*}
    In words, we compose morphisms by wiring the dotted wires together into 
    an \AND gate, and by considering the monoidal product of the dashed 
    wires as the dashed wire of the composition.
    \item The 2-cell compositions and identities are trivial, and defined 
    in the obvious way.
  \end{itemize}
\end{definition}
The reason why we define $\Bkp$ as a bicategory is that 
1-cell composition in $\Bkp$ is not associative. Indeed, $(f;g);h$ and $f;(g;h)$
are different morphisms, as one can see in the figure below:
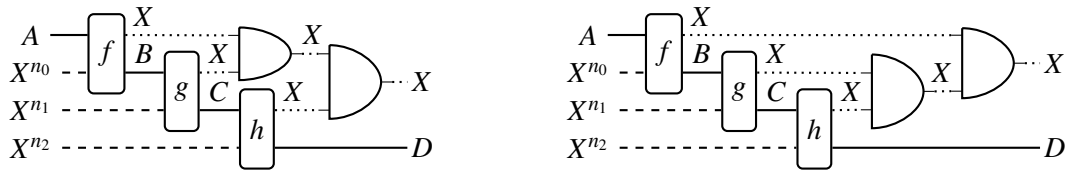
\begin{equation*}
  \begin{tikzpicture}
    \ctikzset{tripoles/american and port/input height=0.7};
    \ctikzset{tripoles/american and port/height=.5};
    \ctikzset{tripoles/american and port/width=.7};
    \node[draw, minimum height=30, minimum width=5, rounded corners=2, thick] (f) at (-1,0.5) {$f$};
    \node[draw, minimum height=30, minimum width=5, rounded corners=2, thick] (g) at (0,0) {$g$};
    \node[draw, minimum height=30, minimum width=5, rounded corners=2, thick] (h) at (1,-0.5) {$h$};
    \node[american and port] (and) at (1.5,0.5) {};
    \ctikzset{tripoles/american and port/input height=0.765};
    \ctikzset{tripoles/american and port/height=.7};
    \ctikzset{tripoles/american and port/width=.7};
    \node[american and port] (and2) at (2.7,0.125) {};

    \node (A) at (-2,0.75) {$A$};
    \node (Xn0) at (-2,0.25) {$X^{n_0}$};
    \node (Xn1) at (-2,-0.25) {$X^{n_1}$};
    \node (Xn2) at (-2,-0.75) {$X^{n_2}$};

    \node (Bf) at (-0.5,1) {$X$};
    \node (B) at (-0.5,0.5) {$B$};

    \node (Bg) at (0.5,0.5) {$X$};
    \node (C) at (0.5,0) {$C$};

    \node (Band) at (1.75,0.75) {$X$};
    \node (Bh) at (1.5,0) {$X$};
    \node (Bfin) at (3.2,0.125) {$X$};
    \node (D) at (3.2,-0.75) {$D$};

    \draw[thick] (A) -- (-1.25,0.75);
    \draw[dashed, thick] (Xn0) -- (-1.25,0.25);
    \draw[dashed, thick] (Xn1) -- (-0.24,-0.25);
    \draw[dashed, thick] (Xn2) -- (0.76,-0.75);

    \draw[dotted, thick] (-0.75,0.75) -- (and.in 1);
    \draw[thick] (-0.75,0.25) -- (-0.24,0.25);

    \draw[dotted, thick] (0.24,0.25) -- (and.in 2);
    \draw[thick] (0.24,-0.25) -- (0.76,-0.25);

    \draw[dotted, thick] (and.out) -- (and2.in 1);
    \draw[dotted, thick] (1.24,-0.25) -- (and2.in 2);

    \draw[dotted, thick] (and2.out) -- (3,0.125);
    \draw[thick] (1.24, -0.75) -- (3,-0.75);
  \end{tikzpicture}
  \qquad\qquad
  \begin{tikzpicture}
    \node[draw, minimum height=30, minimum width=5, rounded corners=2, thick] (f) at (-1,0.5) {$f$};
    \node[draw, minimum height=30, minimum width=5, rounded corners=2, thick] (g) at (0,0) {$g$};
    \node[draw, minimum height=30, minimum width=5, rounded corners=2, thick] (h) at (1,-0.5) {$h$};
    \ctikzset{tripoles/american and port/input height=0.5};
    \ctikzset{tripoles/american and port/height=.7};
    \ctikzset{tripoles/american and port/width=.7};
    \node[american and port] (and) at (2.5,0) {};
    \ctikzset{tripoles/american and port/input height=0.8};
    \ctikzset{tripoles/american and port/height=.665};
    \ctikzset{tripoles/american and port/width=.7};
    \node[american and port] (and2) at (3.7,0.375) {};

    \node (A) at (-2,0.75) {$A$};
    \node (Xn0) at (-2,0.25) {$X^{n_0}$};
    \node (Xn1) at (-2,-0.25) {$X^{n_1}$};
    \node (Xn2) at (-2,-0.75) {$X^{n_2}$};

    \node (Bf) at (-0.5,1) {$X$};
    \node (B) at (-0.5,0.5) {$B$};

    \node (Bg) at (0.5,0.5) {$X$};
    \node (C) at (0.5,0) {$C$};

    \node (Bh) at (1.5,0) {$X$};
    \node (Band) at (2.7,0.25) {$X$};
    \node (Bfin) at (4.2,0.375) {$X$};
    \node (D) at (4.2,-0.75) {$D$};

    \draw[thick] (A) -- (-1.25,0.75);
    \draw[dashed, thick] (Xn0) -- (-1.25,0.25);
    \draw[dashed, thick] (Xn1) -- (-0.24,-0.25);
    \draw[dashed, thick] (Xn2) -- (0.76,-0.75);

    \draw[dotted, thick] (-0.75,0.75) -- (and2.in 1);
    \draw[thick] (-0.75,0.25) -- (-0.24,0.25);

    \draw[dotted, thick] (0.24,0.25) -- (and.in 1);
    \draw[thick] (0.24,-0.25) -- (0.76,-0.25);

    \draw[dotted, thick] (1.24,-0.25) -- (and.in 2);
    \draw[thick] (1.24, -0.75) -- (4,-0.75);

    \draw[dotted, thick] (and.out) -- (and2.in 2);

    \draw[dotted, thick] (and2.out) -- (4,0.375);
  \end{tikzpicture}
\end{equation*}
The point though is that these morphisms implement the same 
boolean function, and are extensionally equal: In fact, it is not difficult 
to check that 
$\texttt{AND}(\texttt{AND}(x,y),z) = \texttt{AND}(x,\texttt{AND}(y,z))$
for each triplet of bits $x, y, z$. A similar argument can be made for identity 
laws, noting that $\texttt{AND}(x, 1) = x = \texttt{AND}(1,x)$ for each 
bit $A$. For these reasons we introduced 
2-cells when $\texttt{ext}f = \texttt{ext}g$, which capture exactly 
the notion of extensional equality. Such cells are by construction invertible 
and give a very trivial 2-structure, where every 2-homset is both 
a preorder and a groupoid, and bicategory axioms hold on the nose.
A more refined definition 
where 2-cells are circuit rewritings could have been given, 
but we are not interested in studying circuit rewriting in this 
work, so we opted for the easiest solution.
\section{Finite State Machines (FSMs)}\label{sec: finite state machines}
We see \emph{state machines} as Petri nets~\cite[Ch.2]{StateboxTeam2019a} where 
each transition has only one inbound and one outbound arc, 
and all markings have exactly one token. In this setting, while 
the usual underlying structure of a Petri net is an hypergraph,
the underlying structure of a state machine is just 
a graph. Another way to put this is that we are freely 
confusing state machines with their state spaces.
\begin{definition}
  A \emph{finite state machine} is a state machine whose
  underlying graph has a finite number of vertexes and edges.
\end{definition}
We can use a Petri net to generate a free symmetric strict monoidal 
category, essentially using its underlying hypergraph structure to define 
object and morphism generators~\cite{Genovese2019}. In the case of FSMs, the restriction 
of their underlying hyergraphs to be graphs simplifies things:
\begin{definition}
  To each FSM $M$ we can assign a \emph{category of executions 
  of $M$}, denoted $\Free{M}$, which is just the free category built 
  on the underlying graph of $M$~\cite[pp.49-51]{MacLane1978}. More in detail, the objects of $\Free{M}$
  are the vertexes of the underlying graph of $M$ (its vertexes), while morphisms 
  are generated by freely composing the edges of the graph. 
  Identities are the null paths. $\Free{-}$ is a functor $\Graph \to \Cat$. 
  It also has a right adjoint, denoted $\UnFree{-}$.
\end{definition}
Given a FSM $M$, every morphism in $\Free{M}$ represents a possible 
run of $M$. The goal for the next section will be to functorially map 
executions into boolean circuits. Then, we will have to turn 
these circuits into boolean circuits, which verify that a given 
execution is correct -- meaning that all the actions performed correspond to a valid path
on the graph. 
\section{Turning executions into circuits}\label{sec: turning executions into circuits}
The first thing to note is that since our graphs are finite, we can 
enumerate their edges and vertexes. We are designing circuits, 
so is important to understand how many bits we need for the 
enumeration. This is seen to be $\lceil \log_2{n} \rceil$, where 
$n$ is the number of elements we need to enumerate. This poses 
another problem: Suppose we have a graph with, say, $6$ 
vertexes. We will need at least $3$ bits to enumerate them. 
Since $2^3 = 8$, we will have two numbers not corresponding 
to any vertex in our enumeration. How do we distinguish 
between numbers enumerating elements and numbers that do 
not? We propose the following solution: First, for each graph $G$ 
with vertexes $V$ and edges $E$, we define functions
$V \to 2^{\lceil \log_2 (|V|+1) \rceil}$ and 
$E \to 2^{\lceil \log_2 (|E+V|) \rceil}$, such that no vertex is 
mapped to $0 \dots 0$ -- the first number of the enumeration, 
from now on also denoted as $\Zero$ -- and no edge is mapped 
to the first $V$ numbers of the enumeration.

The point is that $\Zero$ is reserved in vertex enumerations, 
and is meant to signify \texttt{undefined}. The first $|V|$ numbers 
in the edge enumeration are instead reserved to represent the 
identity morphisms on each vertex in $\Free{G}$. 

Having enumerated vertexes and edges, from the structure of 
the graph we can obtain two tables with the following structure 
template, respectively called \emph{source} and \emph{target table}:
\begin{center}\scriptsize
  \begin{tabular}{c|ccccccccc}
                  & $\Id{v_1}$ & $\cdots$ & $\Id{v_n}$ & $e_1$     & $\cdots$ & $e_m$    & $u_1$     & $\cdots$ & $u_k$   \\
  \hline
  $\Zero$   & 0                 & $\cdots$ & 0                  & 0             & $\cdots$ & 0             & 0             & $\cdots$ & 0             \\
  $v_1$      & 1                 & $\cdots$ & 0                  & ?             & $\cdots$ & ?             & 0             & $\cdots$ & 0             \\
  $\vdots $ & $\vdots$     & $\ddots$ & $\vdots$      & $\vdots$ & $\ddots$ & $\vdots$ & $\vdots$ & $\ddots$ & $\vdots$  \\
  $v_n$      & 0                 & $\cdots$ & 1                  & ?             & $\cdots$ & ?             & 0             & $\cdots$ & 0             \\
  $u'_1$     & 0                 & $\cdots$ & 0                  & 0             & $\cdots$ & 0             & 0             & $\cdots$ & 0             \\
  $\vdots$  & $\vdots$     & $\ddots$ & $\vdots$      & $\vdots$ & $\ddots$ & $\vdots$ & $\vdots$ & $\ddots$ & $\vdots$ \\
  $u'_h$     & 0                 & $\cdots$ & 0                  & 0             & $\cdots$ &0             & 0             & $\cdots$ & 0             \\
  \end{tabular}
\end{center}
Here we have that $n+h+1 = 2^{\lceil \log_2 (|V|+1) \rceil}$, 
and $n+m+k = 2^{\lceil \log_2 (|E+V|) \rceil}$. The $v_i$ are 
the enumerations of the vertexes, the $e_i$ enumerations 
of the edges, and the $u_i, u'_i$ represent the unassigned 
edge and vertex enumerations, respectively. In the source (resp. target) 
table, we put a $1$ in a position if a given vertex is the source (resp. target)
of a given morphism. We reserve the first $n$ enumerations for the 
vertexes for identity morphisms; this forces our 
choices in the first $n$ columns, which along with rows $1$ to $n$ 
define an identity matrix. Similarly, since the $u_i$ and 
$u'_i$ are undefined, there are $0$s in all the entries indexed by them.
The question marks represent the fact that there may be a $0$ or a $1$
in that position, as long as there is just one $1$ in each of those 
columns (an edge can only have one source/target vertex). 
\subsection{Basic circuits}
Using our tables, we are able to build a couple of
boolean functions, where we denoted with $\Bool^V$ and $\Bool^E$
the sets $\Bool^{ \lceil \log_2 (|V|+1) \rceil}$ and 
$\Bool^{\lceil \log_2 (|E+V|) \rceil}$, respectively:
\begin{equation*}
  \Source{G}{\_}, \Target{G}{\_}: \Bool^E \to \Bool^V 
\end{equation*}
These functions take in input the enumeration of an edge, 
and return the enumeration of its source and target vertex, respectively. 
If the input corresponds to an undefined edge, then they return $\Zero$.

The next step is to consider a ``matching function'' 
$\MATCHSym_n: \Bool^n \Tensor \Bool^n \to \Bool$, for each $n$, 
which has the following behaviour:
\begin{equation*}
  \MATCHSym_n (x,y) := \begin{cases}
    1 \text{ iff } (x = y) \wedge (x,y \neq \Zero);\\
    0 \text{ otherwise}.
  \end{cases}
\end{equation*}
Essentially, $\MATCHSym_n$ matches inputs but returns $0$ if one of the inputs
is undefined. We call the boolean circuits implementing $\Source{G}{\_}$,
$\Target{G}{\_}$ and $\MATCHSym_{X^V}(\_, \_)$, $S_G$, $T_G$ and
$\MATCHSym_{X^V}$ respectively.
An example of a circuit implementing 
$\MATCHSym_2$ (so for $2$ bits) together
with its truth table is the following:
\begin{center}
\resizebox{6cm}{2.5cm}{
\begin{circuitikz}[baseline=70pt]
  \draw (-1,2.14)  node[american not port, scale = 0.3] (a1){};
  \draw (-1,3.14)  node[american not port, scale = 0.3] (a2){};
  \draw (-1,3.86)  node[american not port, scale = 0.3] (a3){};
  \draw (-1,4.86)  node[american not port, scale = 0.3] (a4){};

  \draw (0,0)  node[american and port, scale = 0.5] (b1){};
  \draw (0,1) node[american and port, scale = 0.5] (b2){};
  \draw (0,2)  node[american and port, scale = 0.5] (b3){};
  \draw (0,3) node[american and port, scale = 0.5] (b4){};
  \draw (0,4)  node[american and port, scale = 0.5] (b5){};
  \draw (0,5) node[american and port, scale = 0.5] (b6){};

  \draw (1,0.5)  node[american and port, scale = 0.5] (c1){};
  \draw (1,2.5)  node[american and port, scale = 0.5] (c2){};
  \draw (1,4.5)  node[american and port, scale = 0.5] (c3){};

  \draw (2,1.5)  node[american or port, scale = 0.5] (d1){};

  \draw (3,2.5)  node[american or port, scale = 0.5] (e1){};

  \draw[-] (d1.out) |- (e1.in 2);
  \draw[-] (c3.out) |- (e1.in 1);

  \draw[-] (c1.out) |- (d1.in 2);
  \draw[-] (c2.out) |- (d1.in 1);

  \draw[-] (b1.out) |- (c1.in 2);
  \draw[-] (b2.out) |- (c1.in 1);
  \draw[-] (b3.out) |- (c2.in 2);
  \draw[-] (b4.out) |- (c2.in 1);
  \draw[-] (b5.out) |- (c3.in 2);
  \draw[-] (b6.out) |- (c3.in 1);

  \draw[-] (a1.out) |- (b3.in 1);
  \draw[-] (a2.out) |- (b4.in 1);
  \draw[-] (a3.out) |- (b5.in 2);
  \draw[-] (a4.out) |- (b6.in 2);

  \draw[-*] let \p1=(b1.in 2) in (-3,6) to ([yshift=-2pt]-3,\y1);
  \draw[-] let \p1=(b1.in 2) in (-3,\y1) to (b1.in 2);
  \draw[-*] let \p1=(b1.in 1) in (-2.5,6) to ([yshift=-2pt]-2.5,\y1);
  \draw[-] let \p1=(b1.in 1) in (-2.5,\y1) to (b1.in 1);
  \draw[-*] let \p1=(b2.in 2) in (-2,6) to ([yshift=-2pt]-2,\y1);
  \draw[-] let \p1=(b2.in 2) in (-2,\y1) to (b2.in 2);
  \draw[-*] let \p1=(b2.in 1) in (-1.5,6) to ([yshift=-2pt]-1.5,\y1);
  \draw[-] let \p1=(b2.in 1) in (-1.5,\y1) to (b2.in 1);
  
  \draw[-*] let \p1=(a1.in) in (-2.5,6) to ([yshift=-2pt]-2.5,\y1);
  \draw[-] let \p1=(a1.in) in (-2.5,\y1) to (a1.in);
  \draw[-*] let \p1=(b3.in 2) in (-3,6) to ([yshift=-2pt]-3,\y1);
  \draw[-] let \p1=(b3.in 2) in (-3,\y1) to (b3.in 2);
  
  \draw[-*] let \p1=(a2.in) in (-1.5,6) to ([yshift=-2pt]-1.5,\y1);
  \draw[-] let \p1=(a2.in) in (-1.5,\y1) to (a2.in);
  \draw[-*] let \p1=(b4.in 2) in (-2,6) to ([yshift=-2pt]-2,\y1);
  \draw[-] let \p1=(b4.in 2) in (-2,\y1) to (b4.in 2);
  
  \draw[-*] let \p1=(a3.in) in (-3,6) to ([yshift=-2pt]-3,\y1);
  \draw[-] let \p1=(a3.in) in (-3,\y1) to (a3.in);
  \draw[-*] let \p1=(b5.in 1) in (-2.5,6) to ([yshift=-2pt]-2.5,\y1);
  \draw[-] let \p1=(b5.in 1) in (-2.5,\y1) to (b5.in 1);
  \draw[-*] let \p1=(a4.in) in (-2,6) to ([yshift=-2pt]-2,\y1);
  \draw[-] let \p1=(a4.in) in (-2,\y1) to (a4.in);
  \draw[-*] let \p1=(b6.in 1) in (-1.5,6) to ([yshift=-2pt]-1.5,\y1);
  \draw[-] let \p1=(b6.in 1) in (-1.5,\y1) to (b6.in 1);
\end{circuitikz}}
\hspace{2em}
\tiny{\begin{tabular}{c|c|c}
  $a$ & $b$ & $\MATCHSym_2 (a,b)$ \\
  \hline
  $\Zero$   & $\Zero$ & 0 \\
  $01$       & $\Zero$ & 0 \\
  $10$       & $\Zero$ & 0 \\
  $11$       & $\Zero$ & 0 \\
  $\Zero$   & $01$ & 0 \\
  $01$       & $01$ & 1 \\
  $10$      & $01$ & 0 \\
  $11$      & $01$ & 0 \\
  $\Zero$   & $10$ & 0 \\
  $01$       & $10$ & 0 \\
  $10$      & $10$ & 1 \\
  $11$      & $10$ & 0 \\
  $\Zero$   & $11$ & 0 \\
  $01$       & $11$ & 0 \\
  $10$      & $11$ & 0 \\
  $11$      & $11$ & 1 \\
\end{tabular}}
\end{center}
\subsection{Mapping paths}
Denoting with $\COPYSym_{X^E}$ the $\COPY$ circuit acting 
on $E$ bits, we now notice that the boolean circuit 
$(\Id{X^V} \Tensor \COPYSym_{X^E});(\Id{X^V} 
  \Tensor S_G \Tensor T_G);(\MATCHSym_{X^V} \Tensor \Id{X^V})$,
when mapped to $\Bfun$ through $\texttt{ext}$, will correspond to the 
function accepting a vertex and an edge enumeration in input, and 
will return $1$ if the vertex is the source of the edge, $0$ otherwise,
along with the enumeration of the edge's target.
Importantly, it will always return $0$ on the first output for 
any undefined enumeration in input. It is, moreover, a morphism 
in $\Bkp$, as becomes evident by drawing it:
\begin{equation*}
  \begin{tikzpicture}
    \node[draw, circle, radius=5pt, thick] (copy) at (-1,0.5) {};
    \node[draw, minimum height=20, minimum width=20, rounded corners=2, thick] (S) at (0,1) {$S_G$};
    \node[draw, minimum height=20, minimum width=20, rounded corners=2, thick] (T) at (0,0) {$T_G$};
    \node[draw, fill=gray, circle, radius=5pt, thick] (mat) at (1,1.5) {};

    \node[draw, dashed, minimum height=80, minimum width=80] (contour) at (0,0.8) {};

    \node (Xv) at (-2,2) {$X^V$};
    \node (Xe) at (-2,0.5) {$X^E$};
    \node (X) at (2,1.5) {$X$};
    \node (Xv2) at (2,0) {$X^V$};

    \draw[-, thick] (Xv) to (10pt,2);
    \draw[bend left, thick] (10pt,2) to (mat.north);
    \draw[bend right, thick] (S.east) to (mat.south);

    \draw[dashed, -, thick] (Xe) to (copy.west);

    \draw[dashed, bend left, thick] (copy.north) to (S.west);
    \draw[dashed, bend right, thick] (copy.south) to (T.west);

    \draw[-, dotted, thick] (mat.east) -- (X);
    \draw[-, thick] (T.east) -- (Xv2.west);
  \end{tikzpicture}
\end{equation*}
\begin{theorem}\label{thm: path functor}
  Having chosen an enumeration on the vertexes and edges of a graph $G$, there 
  is a pseudofunctor $\Free{G} \to \Bkp$, sending 
  each object to $X^V$, and each generating morphism $e$ of $\Free{G}$ to the 
  following morphism, where $e$ represents the 
  constant gate outputting the enumeration of $e$ when considered as an edge of $G$:
  \begin{equation*}
    (\Id{X^V} \Tensor e);(\Id{X^V} \Tensor \COPYSym_{X^E});
      (\Id{X^V} \Tensor S_G \Tensor T_G);(\MATCHSym_{X^V} \Tensor \Id{X^V})
  \end{equation*}
  \begin{equation*}
    \begin{tikzpicture}
      \node[draw, circle, radius=5pt, thick] (copy) at (-1,0.5) {};
      \node[draw, minimum height=20, minimum width=20, rounded corners=2, thick] (S) at (0,1) {$S_G$};
      \node[draw, minimum height=20, minimum width=20, rounded corners=2, thick] (T) at (0,0) {$T_G$};
      \node[draw, fill=gray, circle, radius=5pt, thick] (mat) at (1,1.5) {};

      \node[draw, dashed, minimum height=80, minimum width=110] (contour) at (-0.5,0.8) {};

      \node (Xv) at (-3,2) {$X^V$};
      \node[draw, regular polygon, regular polygon sides = 3, rotate=-30, thick, label={[rotate=4]center:$e$}] (Xe) at (-1.75,0.5) {};
      \node (X) at (2,1.5) {$X$};
      \node (Xv2) at (2,0) {$X^V$};

      \draw[-, thick] (Xv) to (10pt,2);
      \draw[bend left, thick] (10pt,2) to (mat.north);
      \draw[bend right, thick] (S.east) to (mat.south);

      \draw[dashed, -, thick] (Xe) to (copy.west);

      \draw[dashed, bend left, thick] (copy.north) to (S.west);
      \draw[dashed, bend right, thick] (copy.south) to (T.west);

      \draw[-, dotted, thick] (mat.east) -- (X);
      \draw[-, thick] (T.east) -- (Xv2.west);
    \end{tikzpicture}
  \end{equation*}
  The image of $\Free{G}$ through this pseudofunctor is called 
  $\Bpath{G}$, the category of \emph{path proofs over $G$.}
\end{theorem}
The circuits of Theorem~\ref{thm: path functor} have the disadvantage of 
working on fixed paths, while we would like a general circuit 
working with every path of a given graph. To solve this problem, we take 
an intermediate step:
\begin{lemma}\label{lem: functor to count}
  Consider the category $\Count$, which has one object $*$ and 
  natural numbers as morphisms, with $0$ as the identity morphism 
  and composition as addition. 

  For each graph $G$, there is a functor $\Free{G} \to \Count$ sending 
  every object to  $*$, identities to $0$, and generating morphisms to $1$.
  This extends to a functorial correspondence between $\Graph$ and the 
  category of endofunctors over $\Count$.
\end{lemma}
$\Count$ is a category that, as the name suggests, counts how many 
generating morphisms compose a path. We can use it to shape general 
circuits that work for every path in a graph.
\begin{theorem}\label{thm: graph functor}
  For a graph $G$, consider an enumeration and $S_G$ and $T_G$ 
  as defined in Theorem~\ref{thm: path functor}. There is a pseudofunctor 
  $\Count \to \Bkp$ sending $*$ to $X^V$, $0$ to $\Id{X^V}$ 
  and $n > 0$ to the $n$-fold composition of the morphism
  \begin{equation*}
      (\Id{X^V} \Tensor \COPYSym_{X^E});
      (\Id{X^V} \Tensor S_G \Tensor T_G);(\MATCHSym_{X^V} \Tensor \Id{X^V})
  \end{equation*}
  The composition of this pseudofunctor with the functor of 
  Lemma~\ref{lem: functor to count} gives a pseudofunctor 
  $\Free{G} \to \Count \to \Bkp$ sending each object to $X^V$, and each 
  generating morphism to the circuit:
  \begin{equation*}
    (\Id{X^V} \Tensor \COPYSym_{X^E});
      (\Id{X^V} \Tensor S_G \Tensor T_G);(\MATCHSym_{X^V} \Tensor \Id{X^V})
  \end{equation*}
  \begin{equation*}
    \begin{tikzpicture}
      \node[draw, circle, radius=5pt, thick] (copy) at (-1,0.5) {};
      \node[draw, minimum height=20, minimum width=20, rounded corners=2, thick] (S) at (0,1) {$S_G$};
      \node[draw, minimum height=20, minimum width=20, rounded corners=2, thick] (T) at (0,0) {$T_G$};
      \node[draw, fill=gray, circle, radius=5pt, thick] (mat) at (1,1.5) {};

      \node[draw, dashed, minimum height=80, minimum width=110] (contour) at (-0.5,0.8) {};

      \node (Xv) at (-3,2) {$X^V$};
      \node (Xe) at (-3,0.5) {$X^E$};
      \node (X) at (2,1.5) {$X$};
      \node (Xv2) at (2,0) {$X^V$};

      \draw[-, thick] (Xv) to (10pt,2);
      \draw[bend left, thick] (10pt,2) to (mat.north);
      \draw[bend right, thick] (S.east) to (mat.south);

      \draw[dashed, -, thick] (Xe) to (copy.west);

      \draw[dashed, bend left, thick] (copy.north) to (S.west);
      \draw[dashed, bend right, thick] (copy.south) to (T.west);

      \draw[-, dotted, thick] (mat.east) -- (X);
      \draw[-, thick] (T.east) -- (Xv2.west);
    \end{tikzpicture}
  \end{equation*}
  The image of $\Free{G}$ through this pseudofunctor 
  is called $\Bgraph{G}$, the category of \emph{proofs over $G$.}
\end{theorem}
The pseudofunctor in Theorem~\ref{thm: graph functor} 
associates to each path of length $m$, seen as a morphism 
in $\Free{G}$, a boolean circuit. 
This circuit accepts the enumeration of a vertex $v$ 
and a path of $n$ edges (specified as $n$ enumerations 
of edges) as inputs, and returns $1$ and an enumeration of $v'$
in output if the path leads from $v$ to $v'$. It returns $0$ and 
$\Zero$ otherwise. Notice that since we included identities 
in the truth tables when defining $S_G, T_G$, we are also 
able to process \emph{any path of length less than $n$ by 
padding it with identities}.
\subsection{Snarkizing circuits}\label{subsec: snarkizing circuits}
How do we turn the morphisms in 
$\Bgraph{G}$ into zk-SNARKs? 
Luckily enough, it turns out we do not have to build 
zk-SNARKs ourselves.  Indeed, there are already 
implemented ways to turn boolean circuits into 
zk-SNARKs~\cite{Scipr-lab2012}. Figuring out a cryptographically secure 
way to turn circuits into zk-SNARKs is no simple endeavour, 
that would probably take years and extensive security 
auditing. Instead, we deem a wiser course of action 
turning boolean circuits in 
$\Bgraph{G}$ into boolean circuits, and feed them 
to an already implemented and audited solution.
\begin{definition}\label{def: snarkizator}
  For each graph $G$ we define the \emph{snarkizator} as a function 
  $\Snark{\_} : \Mor{\Bkp} \to \Mor{\Bcirc}$ that 
  maps a morphism $f: A \to B$ to 
  \begin{equation*}
      (f \Tensor \Id{B});(\Id{X} \Tensor \MATCHSym_{B}); \ANDSym
  \end{equation*}
  \begin{equation*}
    \begin{tikzpicture}
      \node[draw, minimum height=30, minimum width=5, rounded corners=2, thick] (f) at (0,1) {$f$};
      \node[draw, fill=gray, circle, radius=5pt, thick] (mat) at (0.75,0.25) {};
      \ctikzset{tripoles/american and port/input height=0.72};
      \ctikzset{tripoles/american and port/height=1};
      \ctikzset{tripoles/american and port/width=1};
      \node[american and port] (and) at (3,0.75) {};

      \node (Xv) at (-2,1.25) {$A$};
      \node (Xe) at (-2,0.75) {$X^n$};
      \node (Xv2) at (-2,-0.25) {$B$};

      \node (X1) at (1.5,1.5) {$X$};
      \node (X2) at (1,0.75) {$B$};
      \node (X2) at (1.5,0.5) {$X$};

      \node (X3) at (4,0.75) {$X$};

      \draw[-, thick] (Xv) to (-0.25,1.25);
      \draw[dashed, -, thick] (Xe) to (-0.25,0.75);
      \draw[-, thick] (Xv2) to (10pt,-0.25);

      \draw[-, dotted, thick] (0.25, 1.25) -- (and.in 1);
      \draw[bend left, thick] (0.25,0.75) to (mat.north);

      \draw[bend right, thick] (0.25,-0.25) to (mat.south);
      \draw[-, dotted, thick] (mat.east) -- (and.in 2);

      \draw[-, dotted, thick] (and.out) -- (X3.west);
    \end{tikzpicture}
  \end{equation*}
\end{definition}
Notice how a snarkized circuit just outputs a bit, which is required to turn it
into a zk-SNARK. Note moreover how the function $\Snark{\_}$ cannot be improved
to a (pseudo)functor, since it does not respect composition.

Since $\Bgraph{G}$ is a subcategory of $\Bkp$, for each
morphism $f$ in $\Bgraph{G}$ we can consider $\Snark{f}$.
It is a boolean circuit which takes 
two values $a,b$ of type $X^V$ as input, representing vertexes, 
along with $f_1, \dots, f_n$ inputs of type $X^E$, representing 
edges, and returns $1$ if and only if the edge inputs define a valid path 
from $a$ to $b$ according to the graph specification defined by $S_G$ and $T_G$.
The corresponding zk-SNARK, obtained by simply feeding our circuit to 
any already available library such as \texttt{libsnark}~\cite{Scipr-lab2012}, is a succint,
non-interactive zero knowledge proof that any specified 
path in the graph -- up to length $n$ -- is valid or not.
\section{Abstracting over graphs}\label{sec: abstracting over graphs}
Up to now, circuits in~$\Bgraph{G}$ have the problem that 
the topology of $G$ is used to define $S_G$ and $T_G$, and is thus 
hardwired in the circuit. Since in creating zk-SNARKs some 
information has to be necessarily made public~\cite{Gennaro2013}, this may cause 
problems. Indeed, it may be possible to reverse-engineer this 
public information to obtain information about $G$. Albeit 
this would still allow to keep used paths secret, the topology 
of the state space of a finite state machine can reveal a lot about 
what the finite state machine is used for. We would like to keep this 
information private.

To do this, we notice that as $S_G$ and $T_G$ are obtained by building 
a truth table from the adjacencies of $G$, this truth 
table could be fed itself to a ``universal function'' that builds $S$ and $T$
for all graphs up to a given size. In detail, if $n, m$ are numbers, 
we can consider boolean functions
\begin{equation*}
  \Source{m,n}{\_}, \Target{m,n}{\_}: \Bool^{f(m,n)} \times \Bool^m \to \Bool^n 
\end{equation*}
Which take $m$ bits in input, representing an enumeration of 
the edges of a graph whose source and target matrices are 
specified by $f(m,n)$ bits, and return $n$ bits, representing
an enumeration of their source and target, respectively. Notice how we 
write $f(m,n)$ since the size of the adjacency matrixes defining 
a graph will in general depend on the maximum number of 
vertexes and edges we are allowing. As before, we consider implementations $S_{m,n}$ 
and $T_{m,n}$ of $\Source{m,n}{\_}$ and $\Target{m,n}{\_}$, 
respectively.

Since we have introduced new inputs, just substituting $S_{m,n}$ 
and $T_{m,n}$ in place of $S_G$ and $T_G$ in 
Theorems~\ref{thm: path functor} and~\ref{thm: graph functor}
won't work: Indeed, it would force us to specify the encoding
for $G$ $k$ times in a $k$-fold composition. Moreover, what happens 
if we give different graphs specifications in the composition?

As we see, building on top of Definition~\ref{def: definition of bkp}
in this new setting creates possibly pathological outcomes. 
The point is that the categorical structure 
of $\Bkp$ is not good to manage inputs that have to be 
routinely repeated. We fix these problems straight away 
by defining a new category as follows:
\begin{definition}\label{def: definition of bzkp}
  Fix a natural number $k$. We denote with $\Bzkp{k}$ 
  the \emph{bicategory of zero knowledge proof circuits of size $k$}, defined 
  as follows:
  \begin{itemize}
    \item $\Obj{\Bzkp{k}} := \Obj{\Bcirc}$;
    \item $\Mor{\Bzkp{k}}(A,B) := \Mor{\Bcirc}(A \Tensor X^k \Tensor X^n, X \Tensor B )$, for all $n \in \Naturals$.
    We depict morphisms as shown below; the $X^k$, $X^n$ and $X$ wires are ``silent'' with respect to 
    our categorical structure, so we depict them densely dotted, dashed and dotted, respectively:
    \begin{equation*}
    \begin{tikzpicture}
      \node[draw, minimum height=40, minimum width=5, rounded corners=2, thick] (f) at (-1,0.5) {$f$};
      \node (A) at (-2,1) {$A$};
      \node (Xk0) at (-2,0.5) {$X^{k}$};
      \node (Xn0) at (-2,0) {$X^{n}$};

      \node (Bf) at (0,1) {$X$};
      \node (B) at (0,0) {$B$};

      \draw[thick] (A) -- (-1.25,1);
      \draw[densely dotted,thick] (Xk0) -- (-1.25,0.5);
      \draw[dashed,thick] (Xn0) -- (-1.25,0);

      \draw[dotted,thick] (-0.75,1) -- (Bf);
      \draw[thick] (-0.75,0) -- (B);
    \end{tikzpicture}  
    \end{equation*}
    \item $\Id{A} := \top \Tensor \Id{A} \Tensor \Id{X^k}: A \Tensor X^k \Tensor X^0  \to X \Tensor A$. Identites are
    depicted as follows:
    \begin{equation*}
    \begin{tikzpicture}
      \node[draw, dashed, minimum height=50, minimum width=20, rounded corners=2, thick] (f) at (-1,0.5) {};
      \node[thick, draw, circle, radius=2pt, inner sep = 2pt] (true) at (-1,1) {\tiny{$\top$}};
      \node (A) at (-2,0.5) {$A$};
      \node (Xk) at (-2,0) {$X^k$};

      \node (Bf) at (0,1) {$X$};
      \node (B) at (0,0.5) {$A$};
      \node (Xk2) at (0,0) {$X^k$};

      \draw[thick] (A) -- (B);
      \draw[densely dotted, thick] (Xk) -- (Xk2);

      \draw[dotted, thick] (-0.75,1) -- (Bf);
    \end{tikzpicture}  
    \end{equation*}
    \item $\Mor{\Bzkp{1}}(A,B)(f,g) = \begin{cases}
      \{*\} & \text{ iff } \texttt{ext}f = \texttt{ext}g;\\
      \,\,\,\, \emptyset & \text{ otherwise}
    \end{cases}$
    \item Given $f : A \to B$ and $g: B \to C$, corresponding to 
    morphisms of $\Bcirc$ $A \Tensor X^{k} \Tensor X^{n_0} \to X \Tensor B$ and
    $B \Tensor X^{k} \Tensor X^{n_1} \to X \Tensor C$, respectively, 
    we set $f;g$ to be the morphism 
    \begin{equation*}
      (\Id{A} \Tensor \COPYSym_{X^k} \Tensor \Id{X^{n_0} \Tensor X^{n_1}});
      (\Id{A \Tensor X^{k}} \Tensor \sigma_{X^{n_0}, X^{k}} \Tensor \Id{X^{n_1}});
      (f \Tensor \Id{X^{k} \Tensor X^{n_1}});(\Id{X} \Tensor g);(\ANDSym \Tensor \Id{C})
    \end{equation*}
    Where we denoted with $\sigma_{X^{n_0}, X^{k_1}}$
    the usual symmetries. Composition is depicted graphically as follows:
    \begin{equation*}
      \begin{tikzpicture}
        \node[draw, circle, radius=1pt, thick] (copy) at (-3,1.25) {};
        \node[draw, minimum height=50, minimum width=5, rounded corners=2, thick] (f) at (-1,1.5) {$f$};
        \node[draw, minimum height=50, minimum width=5, rounded corners=2, thick] (g) at (0,0.5) {$g$};
        \ctikzset{tripoles/american and port/input height=0.7};
        \ctikzset{tripoles/american and port/height=1};
        \ctikzset{tripoles/american and port/width=.7};
        \node[american and port] (and) at (2,1.5) {};

        \node (A) at (-4,2) {$A$};
        \node (Xk) at (-4,1.25) {$X^{k}$};
        \node (Xn0) at (-4,0.5) {$X^{n_0}$};
        \node (Xn1) at (-4,-0) {$X^{n_1}$};

        \node (Bf) at (-0.5,2.25) {$X$};
        \node (B) at (-0.5,1.25) {$B$};
        \node (Bg) at (0.5,1.25) {$X$};
        \node (X) at (3,1.5) {$X$};
        \node (C) at (3,0) {$C$};
        
        \draw[thick] (A) -- (-1.25,2);
        \draw[densely dotted, thick] (Xk) -- (copy.west);
        \draw[dashed, thick] (Xn0) to (-2.5,0.5);
        \draw[dashed, thick] (Xn1) -- (-0.25,0);

        \draw[densely dotted, thick, out=45, in=180] (copy.north east) to (-2.5,1.5);
        \draw[densely dotted, thick, out=-45, in=180] (copy.south east) to (-2.5,1);

        \draw[densely dotted, thick, out=0, in=180] (-2.5,1.5) to (-1.25,1.5);
        \draw[densely dotted, thick, out=0, in=180] (-2.5,1) to (-2,0.5);
        \draw[dashed, thick] (-2.5,0.5) to (-2,1);

        \draw[dashed, thick] (-2,1) -- (-1.25,1);
        \draw[densely dotted, thick] (-2,0.5) -- (-0.25,0.5);

        \draw[dotted, thick] (-0.75,2) -- (and.in 1);
        \draw[thick] (-0.75,1) -- (-0.24, 1);
        \draw[dotted, thick] (0.25,1) -- (and.in 2);

        \draw[dotted, thick] (and.out) -- (X);
        \draw[thick] (0.24,0) -- (C);
      \end{tikzpicture}
    \end{equation*}
    In words, we compose morphisms by wiring the dotted wires together into 
    an \AND gate, by considering the monoidal product of the dashed 
    wires as the dashed wire of the composition, and by feeding a copy the densely 
    dotted input to both circuits.
    \item The 2-cell compositions and identities are trivial, and defined 
    in the obvious way.
  \end{itemize}
\end{definition}%
Proceeding as in Lemma~\ref{lem: functor to count}, 
we are able to prove the following theorem:
\begin{theorem}\label{thm: bzkp functor}
  For each $n,m$, denote with $f(m,n)$ the function outputting how many
  bits are needed to store the source and target truth tables for graphs 
  with $n$ vertexes and $m$ edges. There is a functor $\Count \to \Bzkp{f(m,n)}$ sending each 
  number $k$ to the $k$-fold composition of the morphism
  \begin{equation*}
    (\Id{X^n} \Tensor \COPYSym_{X^{f(m,n)}} \Tensor \COPYSym_{X^m});
    (\Id{X^n \Tensor X^{f(m,n)}} \Tensor \sigma_{X^{f(m,n)},X^m} \Tensor \Id{X^m});
    (\Id{X^n} \Tensor S_{m,n} \Tensor T_{m,n});(\MATCHSym_{X^n} \Tensor \Id{X^n})
  \end{equation*}
  \begin{equation*}
    \begin{tikzpicture}
      \node[draw, circle, radius=5pt, thick] (copy2) at (-2,1.75) {};
      \node[draw, circle, radius=5pt, thick] (copy) at (-2,0.25) {};

      \node[draw, minimum height=40, minimum width=20, rounded corners=2, thick] (S) at (0,1.75) {$S_{m,n}$};
      \node[draw, minimum height=40, minimum width=20, rounded corners=2, thick] (T) at (0,0.25) {$T_{m,n}$};
      \node[draw, fill=gray, circle, radius=5pt, thick] (mat) at (1,2.5) {};

      \node[draw, dashed, minimum height=120, minimum width=110] (contour) at (-0.5,1.25) {};

      \node (Xv) at (-3.5,3) {$X^n$};
      \node (Xf) at (-3.5,1.75) {$X^{f(m,n)}$};
      \node (Xe) at (-3.5,0.25) {$X^m$};

      \node (X) at (2,2.5) {$X$};
      \node (Xv2) at (2,0.25) {$X^n$};

      \draw[-, thick] (Xv) to (10pt,3);
      \draw[-, densely dotted, thick] (Xf) to (copy2.west);
      \draw[-, dashed, thick] (Xe) to (copy.west);

      \draw[densely dotted, bend left, thick] (copy2.north) to (-1.5,2.25);
      \draw[densely dotted, bend right, thick] (copy2.south) to (-1.5, 1.25);
      \draw[dashed, bend left, thick] (copy.north) to (-1.5,0.75);
      \draw[dashed, bend right, thick] (copy.south) to (-1.5, -0.25);

      \draw[-, densely dotted, thick] (-1.5, 2.25) to (-0.47, 2.25);
      \draw[dashed, thick, out=0, in=180] (-1.5,0.75) to (-1, 1.25);
      \draw[densely dotted, thick, out=0, in=180] (-1.5,1.25) to (-1,0.75);
      \draw[-, dashed, thick] (-1.5, -0.25) to (-0.47,-0.25);

      \draw[-, dashed, thick] (-1, 1.25) to (-0.47, 1.25);
      \draw[-, densely dotted, thick] (-1, 0.75) to (-0.47,0.75);

      \draw[bend left, thick] (10pt,3) to (mat.north);
      \draw[bend right, thick] (0.45, 2) to (mat.south);

      \draw[dashed, -, thick] (Xe) to (copy.west);

      \draw[-, dotted, thick] (mat.east) -- (X);
      \draw[-, thick] (T.east) -- (Xv2.west);
    \end{tikzpicture}
  \end{equation*}
\end{theorem}
Notice that Theorem~\ref{thm: bzkp functor} is stronger than 
one would expect: As in Theorem~\ref{thm: graph functor} we obtained
a circuit which not only operated on paths of length $n$, but also on all 
paths of smaller length for a given graph $G$, 
here we have something similar: $f(m,n)$ allows us to
feed  to the circuit the specification of \emph{any} graph
that has \emph{up to} $m$ edges and $n$ vertexes! In this sense, 
fixing $m,n$ amounts to fix some upper bounds for the size 
of the graph, exactly as taking the $k$-fold composition of 
the circuit above amounts to fix some upper bound on the size 
of the path we want to process.

Proceeding as in Section~\ref{subsec: snarkizing circuits}, we can 
define a snarkizator for the category $\Bzkp{f(m,n)}$ by trivially 
adapting Definition~\ref{def: snarkizator}. \emph{A snarkized $k$-fold 
composition of the circuit above verifies if a sequence of $k$ or less 
edges in any graph having at most $m$ edges and $n$ vertexes constitutes 
a valid path in the graph or not.} This is exactly what we wanted: 
A zk-SNARK built on such circuit can be used to succintly proved that 
a given piece of data constitutes a valid path in the state graph of a 
specified finite state machine. In other words, \emph{such zk-SNARK
verifies that the rules specified by a given FSM have been followed}.

We conclude by putting everything together, showing how all the 
constructions we built behave compositionally with respect to each other.
\begin{theorem}\label{thm: the cube}
Let $G$, $G'$ be graphs with $n, n'$ vertices and $m, m'$ 
edges, respectively. Denote with $f(m,n)$ the function 
outputting how many bits are needed to store the source 
and target truth tables for graphs with $n$ vertexes and $m$ edges. 
Then for each morphism $G \to G'$ the following diagram commutes:
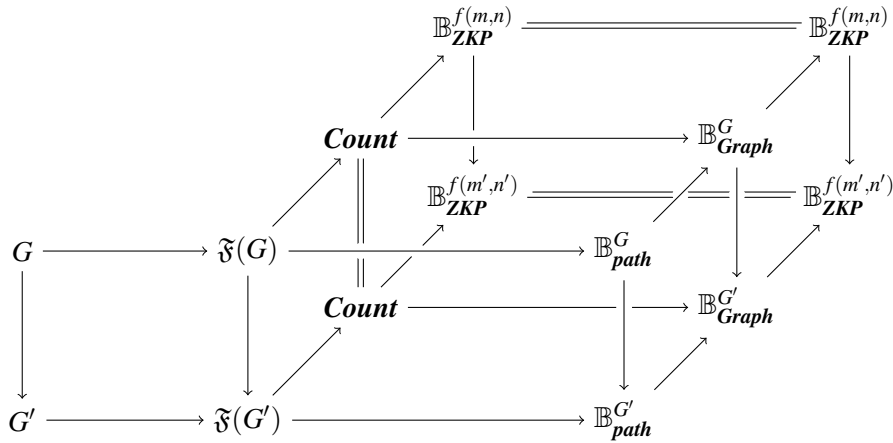
\begin{equation*}
  \begin{tikzpicture}
    \pgfdeclarelayer{fg}
    \pgfdeclarelayer{crossing}
    \pgfdeclarelayer{bg}
    \pgfsetlayers{main,bg,crossing,fg}
    \node (G') at (0,0) {$G'$};
    \node(G) at (0,2.25) {$G$};
    \node(FG') at (3,0) {$\Free{G'}$};
    \node(FG) at (3,2.25) {$\Free{G}$};
    \node(Bpath') at (8,0) {$\Bpath{G'}$};
    \node(Bpath) at (8,2.25) {$\Bpath{G}$};
    \node(Count') at (4.5,1.5) {$\Count$};
    \node(Count) at (4.5,3.75) {$\Count$};
    \node(Bzkp') at (6,3) {$\Bzkp{f(m',n')}$};
    \node(Bzkp) at (6,5.25) {$\Bzkp{f(m,n)}$};
    \node(Bgraph') at (9.5,1.5) {$\Bgraph{G'}$};
    \node(Bgraph) at (9.5,3.75) {$\Bgraph{G}$};
    \node(Bzkp1') at (11,3) {$\Bzkp{f(m',n')}$};
    \node(Bzkp1) at (11,5.25) {$\Bzkp{f(m,n)}$};
      \draw[->] (G) -- (FG);
      \draw[->] (FG) -- (Count);
      \draw[->] (Count) -- (Bzkp);
      \draw[transform canvas={yshift=-1pt, xshift=-1pt}, =] (Bzkp) -- (Bzkp1);
      \draw[transform canvas={yshift=1pt, xshift=1pt}, =] (Bzkp) -- (Bzkp1);
      \draw[->]  (Bgraph) -- (Bzkp1) ;
      \draw[->] (G') -- (FG');
      \draw[->] (FG') -- (Bpath');
      \draw[->] (FG') -- (Count');
      \draw[->] (Bgraph') --  (Bzkp1');
      \draw[->] (Bpath') -- (Bgraph');
      \draw[->] (G) -- (G');
      \draw[->] (FG) -- (FG');
      \draw[->] (Bzkp1) -- (Bzkp1');
    \begin{pgfonlayer}{fg}
      \draw[->, name path global/.expanded=fgbpath] (FG) -- (Bpath);
      \draw[->, name path global/.expanded=countbgraph] (Count) -- (Bgraph);
      \draw[->, name path global/.expanded=bgraphbpath] (Bpath) -- (Bgraph);
      \draw[->, name path global/.expanded = bgraphbgraph'] (Bgraph) -- (Bgraph');
      \draw[->, name path global/.expanded=bpathbpath'] (Bpath) -- (Bpath');
    \end{pgfonlayer}
    \begin{pgfonlayer}{bg}
      \draw[-> , name path global/.expanded=countbzkp'] (Count') -- (Bzkp');
      \draw[->, name path global/.expanded=countbgraph'] (Count') -- (Bgraph');
      \draw[transform canvas={yshift=-1pt, xshift=-1pt}, =, name path global/.expanded=bzkpid1] (Bzkp') -- (Bzkp1');
      \draw[transform canvas={yshift=1pt, xshift=1pt}, =, name path global/.expanded=bzkpid2] (Bzkp') -- (Bzkp1');
      \draw[transform canvas={xshift=-1pt}, =, name path global/.expanded=countcount'1] (Count) -- (Count');
      \draw[transform canvas={xshift=1pt}, =, name path global/.expanded=countcount'2] (Count) -- (Count');
      \draw[->, name path=bzkpbzkp'] (Bzkp) -- (Bzkp');
    \end{pgfonlayer}
    \begin{pgfonlayer}{crossing}
      \fill[white, name intersections={of=bgraphbpath and bzkpid1, name=i}] (i-1) circle (3pt);
      \fill[white, name intersections={of=bgraphbpath and bzkpid2, name=i}] (i-1) circle (3pt);
      \fill[white, name intersections={of=bgraphbgraph' and bzkpid1, name=i}] (i-1) circle (3pt);
      \fill[white, name intersections={of=bgraphbgraph' and bzkpid2, name=i}] (i-1) circle (3pt);
      \fill[white, name intersections={of=fgbpath and countbzkp', name=i}] (i-1) circle (3pt);
      \fill[white, name intersections={of=countbgraph' and bpathbpath', name=i}] (i-1) circle (3pt);
      \fill[white, name intersections={of=bzkpbzkp' and countbgraph, name=i}] (i-1) circle (3pt);
      \fill[white, name intersections={of=countcount'1 and fgbpath, name=i}] (i-1) circle (3pt);
      \fill[white, name intersections={of=countcount'2 and fgbpath, name=i}] (i-1) circle (3pt);
    \end{pgfonlayer}
    \end{tikzpicture}
\end{equation*}
\end{theorem}
\section{Conclusion and future work}\label{sec: conclusion}
We defined a pseudofunctorial way to turn graphs into families of 
boolean circuits that can verify the correctness of any path in the graph.
Then, we generalized this to circuits that can verify correctness of 
paths for any graph with a bounded number of vertexes and edges, 
obtaining a pseudofunctorial correspondence between 
the category $\Graph$ and the category of circuits.

Since graphs can be used to represent finite state machines and 
boolean circuits can be compiled into zk-SNARKs, this in turn 
provides a pseudofunctorial way to turn FSMs into zk-SNARKs, 
with each zk-SNARK verifying that the rules specified by a FSM 
have been followed.

Ongoing work includes implementing our correspondence in a 
formally verified setting using dependent types. To do this, we are 
using \texttt{idris-ct}~\cite{StateboxTeam2019}, our own library 
to do category theory in a dependently typed framework 
(\texttt{Idris}~\cite{Brady2013}).

Future work is 
mainly focused in generalizing our machinery to map free symmetric
strict monoidal categories into boolean circuits, providing a way to 
define circuits verifying executions for Petri nets. Major challenges 
for this task revolve around the fact that the number of tokens in a Petri 
net marking can be \emph{unbounded}. This proves necessary to rethink 
the way we store object-related information in a boolean circuit.

Another interesting line of research revolves around using \emph{recursive 
zk-SNARKs}~\cite{Ben-Sasson2017} to extend the verifying capacities 
of zk-SNARKs beyond a previously fixed upper bound for the edges 
that can compose a graph path. The main idea is that to verify, say, that 
$2n$ edges form a valid path in a graph $G$, we can use a zk-SNARK verifying 
that the first $n$ edges form a path in $G$, and recursively feed it to 
a zk-SNARK verifying that the last $n$ edges form a path in $G$. This recursive 
SNARK then verifies that the overall sequence of $2n$ edges is a valid path in $G$. 
The possibility of recursively composing zk-SNARKs seems very promising to 
generalize our strategy to the verification of paths of arbitrary length.
\section*{Acknowledgements}
The authors want to thank the Ethereum Foundation, 
that financed this work with a grant.

%
%
\section*{Addendum - Building the morphisms $S_G, T_G, S_{m,n}, T_{m,n}$}
In Section~\ref{sec: turning executions into circuits} we mentioned 
a couple of boolean circuits, $S_G, T_G : \Bool^E \to \Bool^V$: 
They depend on a fixed graph $G$, and return 
as output the enumeration of a source and target vertex, respectively, 
of an edge whose enumeration is specified as input.

The beauty of category theory lies precisely in the possibility
 to abstract from the definition of $S_G$ and $T_G$, just focusing 
on how these functions had to be used to assemble the whole verifying 
circuit for $G$. Nevertheless, a precise
description of these circuits is still needed to make an implementation
possible, which we are going to provide in this addendum. We warn the reader 
that our solution is far from being optimal in terms of complexity, so it should be 
understood as a proof of concept.

In the following, we will explain the process for the circuit $S_G$, 
the one for $T_G$ being similar. First, we realize $S_G$ as a table using the incidency 
matrix of the graph $G$:
\begin{center}\scriptsize
  \begin{tabular}{c|c}
    Edge & Source Vertex\\
  \hline
    $\Id{v_1}$   & $v_1$\\
    \dots              & \dots \\
    $\Id{v_n}$   & $v_n$ \\
    $e_1$           & $s(e_1)$ \\
    \dots             & \dots \\
    $e_m$          & $s(e_m)$\\
    $u_1$           & $\Zero$ \\
    \dots             & \dots \\
    $u_k$           & $\Zero$\\
  \end{tabular}
\end{center}
Here $s(\_)$ represents the source function associated to $G$.
From here, we switch to enumerations, where we denote 
with $x^i_j$ the $i$-th binary digit of the enumeration of 
element $x_j$:
\begin{center}\scriptsize
  \begin{tabular}{ccc|ccc}
    Edge 1st bit & & Edge $E$-th bit & Source 1st bit && Source $V$-th bit\\
  \hline
    $\Id{v_1}^1$ & $\dots$ & $\Id{v_1}^E$ & $v_1^1$       & $\dots$ & $v_1^V$\\
                           & $\dots$ &                         &                      & $\dots$ & \\
    $\Id{v_n}^1$ & $\dots$ & $\Id{v_n}^E$ & $v_n^1$       & $\dots$ & $v_n^V$\\
    $e_1^1$          & $\dots$ & $e_1^E$         & $s(e_1)^1$   & $\dots$ & $s(e_1)^V$\\
                            & $\dots$ &                        &                      & $\dots$ & \\
    $e_m^1$         & $\dots$ & $e_m^E$        & $s(e_m)^1$  & $\dots$ & $s(e_m)^V$\\
    $u_1^1$          & $\dots$ & $u_1^E$         & $0$               & $\dots$ & $0$\\
                            & $\dots$ &                        &                      & $\dots$ & \\
    $u_k^1$          & $\dots$ & $u_k^E$         & $0$               & $\dots$ & $0$\\
  \end{tabular}
\end{center}
We now split the table into $V$ different tables, each returning only one digit of the 
vertex enumeration:
\begin{center}\scriptsize
  \begin{tabular}{ccc|c}
    Edge 1st bit & & Edge $E$-th bit & Source 1st bit \\
  \hline
    $\Id{v_1}^1$ & $\dots$ & $\Id{v_1}^E$ & $v_1^1$\\
                           & $\dots$ &                         &\\
    $\Id{v_n}^1$ & $\dots$ & $\Id{v_n}^E$ & $v_n^1$\\
    $e_1^1$          & $\dots$ & $e_1^E$         & $s(e_1)^1$\\
                            & $\dots$ &                        &\\
    $e_m^1$         & $\dots$ & $e_m^E$        & $s(e_m)^1$\\
    $u_1^1$          & $\dots$ & $u_1^E$         & $0$\\
                            & $\dots$ &                        &\\
    $u_k^1$          & $\dots$ & $u_k^E$         & $0$\\
  \end{tabular}
  \hspace{1em}
    $\dots$
  \hspace{1em}
  \begin{tabular}{ccc|c}
    Edge 1st bit & & Edge $E$-th bit & Source $V$-th bit \\
  \hline
    $\Id{v_1}^1$ & $\dots$ & $\Id{v_1}^E$ & $v_1^V$\\
                           & $\dots$ &                         &\\
    $\Id{v_n}^1$ & $\dots$ & $\Id{v_n}^E$ & $v_n^V$\\
    $e_1^1$          & $\dots$ & $e_1^E$         & $s(e_1)^V$\\
                            & $\dots$ &                        &\\
    $e_m^1$         & $\dots$ & $e_m^E$        & $s(e_m)^V$\\
    $u_1^1$          & $\dots$ & $u_1^E$         & $0$\\
                            & $\dots$ &                        &\\
    $u_k^1$          & $\dots$ & $u_k^E$         & $0$\\
  \end{tabular}
\end{center}
Using standard techniques~\cite{Vollmer1999} we are able to turn each one of 
these tables into a circuit. We then obtain a bunch 
of morphisms in $\Bcirc$:
\begin{equation*}
  \begin{tikzpicture}[baseline=0pt]
    \node[draw, minimum height=20, minimum width=20, rounded corners=2, thick] (S) at (0,0) {$S_G^1$};

    \node (Xe) at (-2,0) {$X^E$};
    \node (Xv) at (2,0) {$X$};

    \draw[-, dashed,thick] (Xe) to (S.west);
    \draw[-, dotted, thick] (S.east) to (Xv);
  \end{tikzpicture}
  \qquad \dots \qquad
  \begin{tikzpicture}[baseline=-1pt]
    \node[draw, minimum height=20, minimum width=20, rounded corners=2, thick] (S) at (0,0) {$S_G^V$};

    \node (Xe) at (-2,0) {$X^E$};
    \node (Xv) at (2,0) {$X$};

    \draw[-, dashed, thick] (Xe) to (S.west);
    \draw[-, dotted, thick] (S.east) to (Xv);
  \end{tikzpicture}
\end{equation*}
The resulting morphism $S_G$ is obtained by just copying the 
input and tensoring the morphisms together:
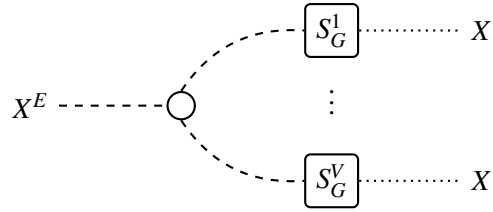
\begin{equation*}
  \begin{tikzpicture}
    \node at (4.25,0) {};
    \node[draw, minimum height=20, minimum width=20, rounded corners=2, thick] (SV) at (0,-1) {$S_G^V$};
    \node[thick] (Sdots) at (0,0.15)  {$\vdots$};
    \node[draw, minimum height=20, minimum width=20, rounded corners=2, thick] (S1) at (0,1)  {$S_G^1$};
    \node[draw, circle, radius=5pt, thick] (copy) at (-2,0) {};

    \node (XeV) at (-4,0) {$X^E$};

    \node (XvV) at (2,-1) {$X$};
    \node (Xv1) at (2,1) {$X$};

    \draw[-, dashed, thick] (XeV) to (copy.west);
    \draw[bend left, dashed, thick] (copy.north) to (S1.west);
    \draw[bend right, dashed, thick] (copy.south) to (SV.west);

    \draw[-, dotted, thick] (SV.east) to (XvV);
    \draw[-, dotted, thick] (S1.east) to (Xv1);
  \end{tikzpicture}
\end{equation*}
Here one should note that the $V$-fold monoidal product of $X$ is 
just $X^V$, so the string diagram above is a morphism of the right type, from 
$X^E$ to $X^V$.

Now we focus on the circuits $S_{m,n}, T_{m,n}: X^{f(m,n)} \otimes X^m \to X^n $.
These accept a graph enumeration of $f(m,n)$ bits in input, along with an edge 
enumeration of $m$ bits, and return a vertex enumeration of $n$ bits, representing the 
source and target, respectively, of the edge provided in the specified graph.

Again, we focus on $S_{m,n}$, the procedure for $T_{m,n}$ being
similar. We start by noticing that for a fixed
bitstring $s$ of length $f(m,n)$, we can produce a circuit 
$F_s: X^{f(m,n)} \to X$ (called \emph{filter for $s$}) that outputs $1$ only 
if the input is $s$, and $0$ otherwise: For each digit composing 
$s$, we concatenate it with a $\NOT$ gate if it is $0$, and we leave it as it is 
otherwise. Then we wire all the bits with $\AND$ ports. For instance, 
the circuit below, working for inputs of $4$ bits, 
outputs $1$ only if the input is $1001$.
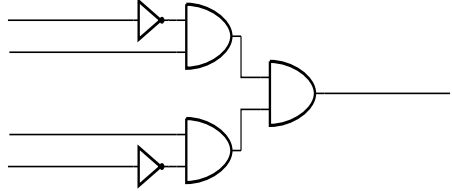
\begin{center}
  \resizebox{6cm}{2.5cm}{
  \begin{circuitikz}
    \draw (-1,-0.14)  node[american not port, scale = 0.3] (a1){};
    \draw (-1,1.14)  node[american not port, scale = 0.3] (a2){};
    \draw (0,0)  node[american and port, scale = 0.5] (b1){};
    \draw (0,1) node[american and port, scale = 0.5] (b2){};
    \draw (1,0.5)  node[american and port, scale = 0.5] (c1){};

    \draw[-] (b1.out) |- (c1.in 2);
    \draw[-] (b2.out) |- (c1.in 1);
    \draw[-] (a2.out) |- (b2.in 1);
    \draw[-] (a1.out) |- (b1.in 2);

    \draw[-] (a1.in) -- +(-1.5,0);
    \draw[-] (a2.in) -- +(-1.5,0);
    \draw[-] (b1.in 1) -- +(-2,0);
    \draw[-] (b2.in 2) -- +(-2,0);

    \draw[-] (c1.out) -- +(1.5,0);
  \end{circuitikz}}
\end{center}
Interpreting a bitstring $s$ as the enumeration of a 
graph $G$ with $m$ edges and $n$ vertexes, we can consider the circuit:
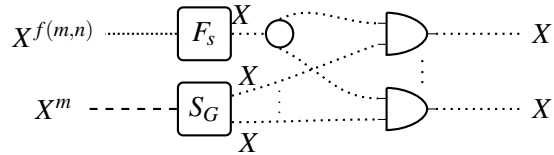
\begin{equation*}
  \begin{tikzpicture}
    \node at (4.25,0) {};
    \node[draw, minimum height=20, minimum width=20, rounded corners=2, thick] (SG) at (0,0) {$S_G$};
    \node[draw, minimum height=20, minimum width=20, rounded corners=2, thick] (F) at (0,1)  {$F_s$};
    \node[draw, circle, radius=5pt, thick] (copy) at (1,1) {};
    \draw (3,0)  node[american and port, scale = 0.5] (a2){};
    \draw (3,1)  node[american and port, scale = 0.5] (a1){};

    \node (XeV) at (-2,0) {$X^m$};
    \node (XF) at (-2,1) {$X^{f(m,n)}$};
    \node (XOut1) at (4.5,1) {$X$};
    \node (XOut2) at (4.5,0) {$X$};

    \node (XSG1) at (0.6,-0.45) {$X$};
    \node (XSGDots) at (1,0.2) {\tiny{$\vdots$}};
    \node (AndDots) at (2.9,0.6) {\scriptsize{$\vdots$}};
    \node (XSG2) at (0.6,0.45) {$X$};
    \node (XF1) at (0.5,1.25) {$X$};
  
    \draw[-, dashed, thick] (SG.west) to (XeV);
    \draw[-, densely dotted, thick] (F.west) to (XF);

    \draw[-, dotted, thick] (F.east) to (copy.west);

    \draw[-, out=25, in=180, dotted, thick] (copy.north) to (a1.in 1);
    \draw[-, out=-25, in=180, dotted, thick] (copy.south) to (a2.in 1);

    \draw[-, out=15, in=200, dotted, thick] ([yshift=5pt]SG.east) to (a1.in 2);
    \draw[-, out=0, in=180, dotted, thick] ([yshift=-5pt]SG.east) to (a2.in 2);

    \draw[-, dotted, thick] (a1.out) to (XOut1.west);
    \draw[-, dotted, thick] (a2.out) to (XOut2.west);
  \end{tikzpicture}
\end{equation*}
That is, we are connecting the output of $F_s$ with each output bit of 
$S_G$ using $\AND$ ports. The result is a circuits 
$S_{m,n}^G: X^{f(m,n)} \otimes X^m \to X^n$ that reduces to $S_G$
when the input on $X^{f(n,m)}$ is the enumeration of the graph $G$, 
while it reduces to the constant $0$ circuit in any other case.

The circuit $S_{m,n}$ is obtained by tensoring together all the 
$S_{m,n}^G$ gates for each graph $G$ of $m$ edges and $v$ vertexes,
by copying the inputs and taking the $\OR$ of the outputs.
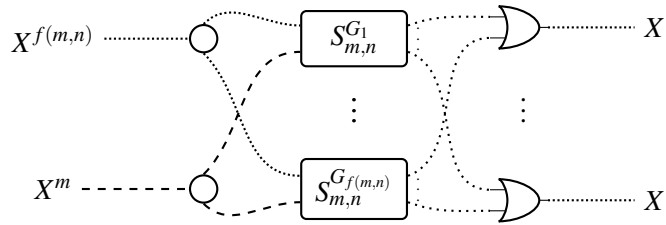
\begin{equation*}
  \begin{tikzpicture}
    \node at (4.25,0) {};
    \node[draw, minimum height=20, minimum width=40, rounded corners=2, thick] (SV) at (0,-1) {$S_{m,n}^{G_{f(m,n)}}$};
    \node[thick] (Sdots) at (0,0.15)  {$\vdots$};
    \node[draw, minimum height=20, minimum width=40, rounded corners=2, thick] (S1) at (0,1)  {$S_{m,n}^{G_1}$};
    \node[draw, circle, radius=5pt, thick] (copyGraph) at (-2,1) {};
    \node[draw, circle, radius=5pt, thick] (copyEdge) at (-2,-1) {};
    \draw (2.5,-1.15)  node[american or port, scale = 0.5] (o2){};
    \draw (2.5,1.15)  node[american or port, scale = 0.5] (o1){};

    \node (Xg) at (-4,1) {$X^{f(m,n)}$};
    \node (Xe) at (-4,-1) {$X^m$};

    \node[thick] (Sdots) at (0,0.15)  {$\vdots$};
    \node[thick] (Ordots) at (2.25,0.15)  {$\vdots$};
    \node (FirstDots) at (0.85,1.1) {\tiny{$\vdots$}};
    \node (SecondDots) at (0.85,-0.9) {\tiny{$\vdots$}};

    \node (XvV) at (4,-1.15) {$X$};
    \node (Xv1) at (4,1.15) {$X$};

    \draw[-, densely dotted, thick] (Xg) to (copyGraph.west);
    \draw[-, dashed, thick] (Xe) to (copyEdge.west);

    \draw[-, in=180, out=45, densely dotted, thick] (copyGraph.north) to ([yshift=5pt]S1.west);
    \draw[-, in=180, out=-45, densely dotted, thick] (copyGraph.south) to ([yshift=5pt]SV.west);
    \draw[-, in=180, out=45, dashed, thick] (copyEdge.north) to ([yshift=-5pt]S1.west);
    \draw[-, in=180, out=-45, dashed, thick] (copyEdge.south) to ([yshift=-5pt]SV.west);

    \draw[-, in=180, out=20, dotted, thick] ([yshift=5pt]SV.east) to (o1.in 2);
    \draw[-, in=180, out=-20, dotted, thick] ([yshift=-5pt]SV.east) to (o2.in 2);

    \draw[-, in=180, out=20, dotted, thick] ([yshift=5pt]S1.east) to (o1.in 1);
    \draw[-, in=180, out=-20, dotted, thick] ([yshift=-5pt]S1.east) to (o2.in 1);

    \draw[-, densely dotted, thick] (o1.out) to (Xv1.west);
    \draw[-, densely dotted, thick] (o2.out) to (XvV.west);
  \end{tikzpicture}
\end{equation*}
Feeding the enumeration of a graph $G$ to the circuit above has the following effect:
Exactly one of the tensored circuits will reduce to $S_G$, while
all the others will reduce to the constant $0$ circuit. Since $0$ is a unit for $\OR$, 
the resulting circuit reduces to $S_G$.
\clearpage
\bibliographystyle{eptcs}
\bibliography{Bibliography}
\appendix
\section*{Appendix - Proofs}
\begingroup
\def\thetheoremUnified{\ref{lem: monoidal functor Bcirc Bfun}}
\begin{theorem}
  There exists a strict monoidal functor $\texttt{ext}^G: \Bcirc^G \to \Bfun$
  sending the generating morphisms $m_g$ to the function $\texttt{int}(g)$.
\end{theorem}
\addtocounter{theoremUnified}{-1}
\endgroup
\begin{proof}
  Strict monoidal functoriality is obvious from the freeness of $\Bcirc$.
\end{proof}
\begin{lemma}\label{lem: bkp is a bicategory}
  $\Bkp$ as defined in Definition~\ref{def: definition of bkp} is a bicategory.
\end{lemma}
\begin{proof}
  First notice that for each $A,B$, $\Bkp(A,B)$ is trivially a category, 
  since function extensionality is reflexive and transitive.

  Moreover, composition is clearly functorial. This follows from the fact 
  that extensionality is a congruence with respect to function composition.

  The existence of unitors and associators follows from the fact that 
  $\ANDSym(x, 1)$ and $\ANDSym(1,x)$ are extensionally equal to the 
  identity function $\Bool \to \Bool$, as extensionally equal are 
  $\ANDSym(\ANDSym(\_,\_),\_)$ and $\ANDSym(\_,\ANDSym(\_,\_))$.

  Naturality and satisfaction of pentagon and triangle identities for 
  associators and unitors, respectively, follows from the 
  fact that the 2-cell structure of $\Bkp$ is a preorder, so proving 
  that such morphisms exist is enough.
\end{proof}
\begingroup
\def\thetheoremUnified{\ref{thm: path functor}}
\begin{theorem}
  Having chosen an enumeration on the vertexes and edges of a graph $G$, there 
  is a pseudofunctor $\Free{G} \to \Bkp$, sending 
  each object to $X^V$, and each generating morphism $e$ of $\Free{G}$ to the 
  following morphism, called \emph{$e$-evaluator}, where $e$ represents the 
  constant gate outputting the enumeration of $e$ when considered as an edge of $G$:
  \begin{equation*}
    (\Id{X^V} \Tensor e);(\Id{X^V} \Tensor \COPYSym_{X^E});
      (\Id{X^V} \Tensor S_G \Tensor T_G);(\MATCHSym_{X^V} \Tensor \Id{X^V})
  \end{equation*}
  \begin{equation*}
    \begin{tikzpicture}
      \node[draw, circle, radius=5pt, thick] (copy) at (-1,0.5) {};
      \node[draw, minimum height=20, minimum width=20, rounded corners=2, thick] (S) at (0,1) {$S_G$};
      \node[draw, minimum height=20, minimum width=20, rounded corners=2, thick] (T) at (0,0) {$T_G$};
      \node[draw, fill=gray, circle, radius=5pt, thick] (mat) at (1,1.5) {};

      \node[draw, dashed, minimum height=80, minimum width=110] (contour) at (-0.5,0.8) {};

      \node (Xv) at (-3,2) {$X^V$};
      \node[draw, regular polygon, regular polygon sides = 3, rotate=-30, thick, label={[rotate=4]center:$e$}] (Xe) at (-1.75,0.5) {};
      \node (X) at (2,1.5) {$X$};
      \node (Xv2) at (2,0) {$X^V$};

      \draw[-, thick] (Xv) to (10pt,2);
      \draw[bend left, thick] (10pt,2) to (mat.north);
      \draw[bend right, thick] (S.east) to (mat.south);

      \draw[dashed, -, thick] (Xe) to (copy.west);

      \draw[dashed, bend left, thick] (copy.north) to (S.west);
      \draw[dashed, bend right, thick] (copy.south) to (T.west);

      \draw[-, dotted, thick] (mat.east) -- (X);
      \draw[-, thick] (T.east) -- (Xv2.west);
    \end{tikzpicture}
  \end{equation*}
  The image of $\Free{G}$ through this pseudofunctor is called 
  $\Bpath{G}$, the category of \emph{path proofs over $G$.}
\end{theorem}
\addtocounter{theoremUnified}{-1}
\endgroup
\begin{proof}
  Obvious from the freeness of $\Free{G}$ and the fact that the 
  bicategorical structure of $\Bkp$ is trivial.
\end{proof}
\begingroup
\def\thetheoremUnified{\ref{thm: path functor}}
\begin{lemma}
  Consider the category $\Count$, which has one object $*$ and 
  natural numbers as morphisms, with $0$ is the identity morphism 
  and composition as addition. 

  For each graph $G$, there is a pseudofunctor $\Free{G} \to \Count$ sending 
  every object to  $*$, identities to $0$, and generating morphisms to $1$.
  This extends to a functorial correspondence between $\Graph$ and the 
  category of endofunctors over $\Count$.
\end{lemma}
\addtocounter{theoremUnified}{-1}
\endgroup
\begin{proof}
  Functoriality is obvious from the freeness of $\Free{G}$.
  Moreover, observe how any morphism of graphs induces a functor between 
  their corresponding free categories which sends generating morphisms 
  to generating morphisms. In particular, this means that such 
  functor preserves the length of paths, making the following diagram 
  trivially commute:
  \begin{equation*}
    \begin{tikzpicture}
      \node (G') at (-2,0) {$G'$}; 
      \node (G) at (-2,2) {$G$};
      \node (FG') at (0,0) {$\Free{G'}$}; 
      \node (FG) at (0,2) {$\Free{G}$};
      \node (Count') at (2,0) {$\Count$};
      \node (Count) at (2,2) {$\Count$};
      \draw[->] (G) -- node[left] {$f$} (G');
      \draw[->] (G) -- (FG);
      \draw[->] (G') -- (FG');
      \draw[->] (FG) -- node[left] {$\Free{f}$} (FG');
      \draw[->] (FG) -- (Count);
      \draw[->] (FG') -- (Count');
      \draw[transform canvas={xshift=-1pt}, -] (Count) -- (Count');
      \draw[transform canvas={xshift=1pt}, -] (Count) -- (Count');
    \end{tikzpicture}
  \end{equation*}
  This suffices to prove the existence of a functorial correspondence $\Graph \to \Cat(\Count, \Count)$.
\end{proof}
\begingroup
\def\thetheoremUnified{\ref{thm: graph functor}}
\begin{theorem}
  For a graph $G$, consider an enumeration and $S_G$ and $T_G$ 
  as defined in Theorem~\ref{thm: path functor}. There is a pseudofunctor 
  $\Count \to \Bkp$ sending $*$ to $X^V$, $0$ to $\Id{X^V}$ 
  and $n > 0$ to the $n$-fold composition of the morphism
  \begin{equation*}
      (\Id{X^V} \Tensor \COPYSym_{X^E});
      (\Id{X^V} \Tensor S_G \Tensor T_G);(\MATCHSym_{X^V} \Tensor \Id{X^V})
  \end{equation*}
  The composition of this pseudofunctor with the pseudofunctor of 
  Lemma~\ref{lem: functor to count} gives a pseudofunctor 
  $\Free{G} \to \Count \to \Bkp$ sending each object to $X^V$, and each 
  generating morphism to the circuit:
  \begin{equation*}
    (\Id{X^V} \Tensor \COPYSym_{X^E});
      (\Id{X^V} \Tensor S_G \Tensor T_G);(\MATCHSym_{X^V} \Tensor \Id{X^V})
  \end{equation*}
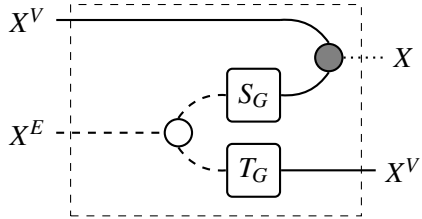
  \begin{equation*}
    \begin{tikzpicture}
      \node[draw, circle, radius=5pt, thick] (copy) at (-1,0.5) {};
      \node[draw, minimum height=20, minimum width=20, rounded corners=2, thick] (S) at (0,1) {$S_G$};
      \node[draw, minimum height=20, minimum width=20, rounded corners=2, thick] (T) at (0,0) {$T_G$};
      \node[draw, fill=gray, circle, radius=5pt, thick] (mat) at (1,1.5) {};

      \node[draw, dashed, minimum height=80, minimum width=110] (contour) at (-0.5,0.8) {};

      \node (Xv) at (-3,2) {$X^V$};
      \node (Xe) at (-3,0.5) {$X^E$};
      \node (X) at (2,1.5) {$X$};
      \node (Xv2) at (2,0) {$X^V$};

      \draw[-, thick] (Xv) to (10pt,2);
      \draw[bend left, thick] (10pt,2) to (mat.north);
      \draw[bend right, thick] (S.east) to (mat.south);

      \draw[dashed, -, thick] (Xe) to (copy.west);

      \draw[dashed, bend left, thick] (copy.north) to (S.west);
      \draw[dashed, bend right, thick] (copy.south) to (T.west);

      \draw[-, dotted, thick] (mat.east) -- (X);
      \draw[-, thick] (T.east) -- (Xv2.west);
    \end{tikzpicture}
  \end{equation*}
  The image of $\Free{G}$ through this pseudofunctor 
  is called $\Bgraph{G}$, the category of \emph{proofs over $G$.}
\end{theorem}
\addtocounter{theoremUnified}{-1}
\endgroup
\begin{proof}
  The only non-trivial part is proving pseudofunctoriality 
  of the functor $\Count \to \Bkp$. For this, notice that 
  the same natural number $n$ can be obtained by 
  adding smaller numbers in different orders, and with 
  different bracketings. These will in turn correspond to 
  different ways to compose the same morphism in $\Bkp$,
  $n$-times. All these compositions are extensionally equal, 
  which guarantees the existence of isomorphic 2-cells between 
  them. Pseudofunctoriality follows trivially, leveraging on the 
  fact that the 2-cell structure of $\Bkp$ is itself trivial.
\end{proof}
\begin{lemma}
  For each $n$, $\Bzkp{n}$ as defined in Definition~\ref{def: definition of bzkp} is a bicategory.
\end{lemma}
\begin{proof}
  The proof proceeds exactly as in Lemma~\ref{lem: bkp is a bicategory}, 
  by applying function extensionality also to different 
  compositions of $\COPYSym_X^m$  and $\sigma_{X^{f(n,m)},X^m}$.
\end{proof}
\begingroup
\def\thetheoremUnified{\ref{thm: the cube}}
\begin{theorem}
  Let $G$, $G'$ be graphs with $n, n'$ vertices and $m, m'$ 
  edges, respectively. Denote with $f(n,m)$ the function 
  outputting how many bits are needed to store the source 
  and target truth tables for graphs with $n$ vertexes and $m$ edges. 
  Then for each morphism $G \to G'$ the following diagram commutes:
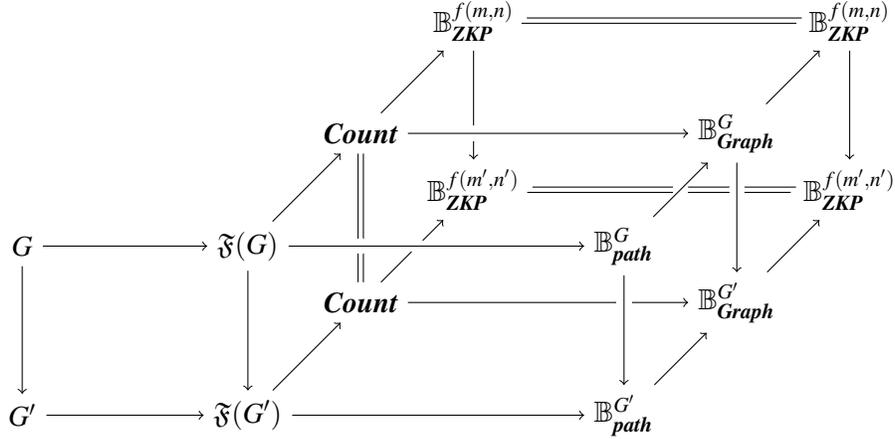
\begin{equation*}
  \begin{tikzpicture}
    \pgfdeclarelayer{fg}
    \pgfdeclarelayer{crossing}
    \pgfdeclarelayer{bg}
    \pgfsetlayers{main,bg,crossing,fg}
    \node (G') at (0,0) {$G'$};
    \node(G) at (0,2.25) {$G$};
    \node(FG') at (3,0) {$\Free{G'}$};
    \node(FG) at (3,2.25) {$\Free{G}$};
    \node(Bpath') at (8,0) {$\Bpath{G'}$};
    \node(Bpath) at (8,2.25) {$\Bpath{G}$};
    \node(Count') at (4.5,1.5) {$\Count$};
    \node(Count) at (4.5,3.75) {$\Count$};
    \node(Bzkp') at (6,3) {$\Bzkp{f(m',n')}$};
    \node(Bzkp) at (6,5.25) {$\Bzkp{f(m,n)}$};
    \node(Bgraph') at (9.5,1.5) {$\Bgraph{G'}$};
    \node(Bgraph) at (9.5,3.75) {$\Bgraph{G}$};
    \node(Bzkp1') at (11,3) {$\Bzkp{f(m',n')}$};
    \node(Bzkp1) at (11,5.25) {$\Bzkp{f(m,n)}$};
      \draw[->] (G) -- (FG);
      \draw[->] (FG) -- (Count);
      \draw[->] (Count) -- (Bzkp);
      \draw[transform canvas={yshift=-1pt, xshift=-1pt}, =] (Bzkp) -- (Bzkp1);
      \draw[transform canvas={yshift=1pt, xshift=1pt}, =] (Bzkp) -- (Bzkp1);
      \draw[->]  (Bgraph) -- (Bzkp1) ;
      \draw[->] (G') -- (FG');
      \draw[->] (FG') -- (Bpath');
      \draw[->] (FG') -- (Count');
      \draw[->] (Bgraph') --  (Bzkp1');
      \draw[->] (Bpath') -- (Bgraph');
      \draw[->] (G) -- (G');
      \draw[->] (FG) -- (FG');
      \draw[->] (Bzkp1) -- (Bzkp1');
    \begin{pgfonlayer}{fg}
      \draw[->, name path global/.expanded=fgbpath] (FG) -- (Bpath);
      \draw[->, name path global/.expanded=countbgraph] (Count) -- (Bgraph);
      \draw[->, name path global/.expanded=bgraphbpath] (Bpath) -- (Bgraph);
      \draw[->, name path global/.expanded = bgraphbgraph'] (Bgraph) -- (Bgraph');
      \draw[->, name path global/.expanded=bpathbpath'] (Bpath) -- (Bpath');
    \end{pgfonlayer}
    \begin{pgfonlayer}{bg}
      \draw[-> , name path global/.expanded=countbzkp'] (Count') -- (Bzkp');
      \draw[->, name path global/.expanded=countbgraph'] (Count') -- (Bgraph');
      \draw[transform canvas={yshift=-1pt, xshift=-1pt}, =, name path global/.expanded=bzkpid1] (Bzkp') -- (Bzkp1');
      \draw[transform canvas={yshift=1pt, xshift=1pt}, =, name path global/.expanded=bzkpid2] (Bzkp') -- (Bzkp1');
      \draw[transform canvas={xshift=-1pt}, =, name path global/.expanded=countcount'1] (Count) -- (Count');
      \draw[transform canvas={xshift=1pt}, =, name path global/.expanded=countcount'2] (Count) -- (Count');
      \draw[->, name path=bzkpbzkp'] (Bzkp) -- (Bzkp');
    \end{pgfonlayer}
    \begin{pgfonlayer}{crossing}
      \fill[white, name intersections={of=bgraphbpath and bzkpid1, name=i}] (i-1) circle (3pt);
      \fill[white, name intersections={of=bgraphbpath and bzkpid2, name=i}] (i-1) circle (3pt);
      \fill[white, name intersections={of=bgraphbgraph' and bzkpid1, name=i}] (i-1) circle (3pt);
      \fill[white, name intersections={of=bgraphbgraph' and bzkpid2, name=i}] (i-1) circle (3pt);
      \fill[white, name intersections={of=fgbpath and countbzkp', name=i}] (i-1) circle (3pt);
      \fill[white, name intersections={of=countbgraph' and bpathbpath', name=i}] (i-1) circle (3pt);
      \fill[white, name intersections={of=bzkpbzkp' and countbgraph, name=i}] (i-1) circle (3pt);
      \fill[white, name intersections={of=countcount'1 and fgbpath, name=i}] (i-1) circle (3pt);
      \fill[white, name intersections={of=countcount'2 and fgbpath, name=i}] (i-1) circle (3pt);
    \end{pgfonlayer}
    \end{tikzpicture}
\end{equation*}
\end{theorem}
\addtocounter{theoremUnified}{-1}
\endgroup
\begin{proof}
  We will prove the commutativity of single squares, 
  starting from the top face. In the square:
  \begin{equation*}
    \begin{tikzpicture}
      \node (FG) at (0,0) {$\Free{G}$}; 
      \node (Count) at (0,2) {$\Count$};
      \node (BpathG) at (2,0) {$\Bpath{G}$}; 
      \node (BgraphG) at (2,2) {$\Bgraph{G}$};
      \draw[->] (FG) -- (Count);
      \draw[->] (Count) -- (BgraphG);
      \draw[->] (FG) -- (BpathG);
      \draw[->] (BpathG) -- (BgraphG);
    \end{tikzpicture}
  \end{equation*}
  We notice that the bottom functor acts by sending 
  each edge $e$ to the morphism:
  \begin{equation*}
    (\Id{X^n} \Tensor e);(\Id{X^n} \Tensor \COPYSym_{X^m});
    (\Id{X^n} \Tensor S_G \Tensor T_G);(\MATCHSym_{X^n} \Tensor \Id{X^n})
  \end{equation*}
  Being a subcategory of $\Bcirc$, which is free symmetric monoidal,
  eavery morphism in $\Bpath{G}$ is just a composition of a finite 
  number of morphisms as the one above for some edges $e_1, \dots, e_n$.
  We then define a pseudofunctor sending each morphism 
  \begin{equation*}
    (\Id{X^n} \Tensor e);(\Id{X^n} \Tensor \COPYSym_{X^m});
    (\Id{X^n} \Tensor S_G \Tensor T_G);(\MATCHSym_{X^n} \Tensor \Id{X^n})
  \end{equation*}
  To:
  \begin{equation*}
    (\Id{X^n} \Tensor \COPYSym_{X^m});
    (\Id{X^n} \Tensor S_G \Tensor T_G);(\MATCHSym_{X^n} \Tensor \Id{X^n})
  \end{equation*}
  The mapping on 2-cells is trivial. Pseudofunctoriality 
  is obvious and follows from function extensionality
  being a congruence wrt function composition. Commutativity 
  is obvious too since the morphism above is precisely where
  precisely where each generating morphism in $\Free{G}$ gets sent to
  when going through $\Count$.

  The square:
  \begin{equation*}
    \begin{tikzpicture}
      \node (Count) at (0,0) {$\Count$};
      \node (BgraphG) at (2,0) {$\Bgraph{G}$};
      \node (BpathG) at (2,2) {$\Bzkp{f(n,m)}$}; 
      \node (FG) at (0,2) {$\Bzkp{f(n,m)}$}; 

      \draw[<-] (FG) -- (Count);
      \draw[->] (Count) -- (BgraphG);
      \draw[transform canvas={yshift=-1pt}, =] (FG) -- (BpathG);
      \draw[transform canvas={yshift=1pt}, =] (FG) -- (BpathG);
      \draw[<-] (BpathG) -- (BgraphG);
    \end{tikzpicture}
  \end{equation*}
  Commutes by noticing that $\Bgraph{G}$ is again generated by the morphism 
  \begin{equation*}
    (\Id{X^n} \Tensor \COPYSym_{X^m});
    (\Id{X^n} \Tensor S_G \Tensor T_G);(\MATCHSym_{X^n} \Tensor \Id{X^n})
  \end{equation*}
   We can then define the right-edge pseudofunctor to be sending it to:
  \begin{equation*}
    (\Id{X^n} \Tensor \COPYSym_{X^{f(n,m)}} \Tensor \COPYSym_{X^m});
    (\Id{X^n \Tensor X^{f(n,m)}} \Tensor \sigma_{X^{f(n,m)},X^m} \Tensor \Id{X^m});
    (\Id{X^n} \Tensor S_{n,m} \Tensor T_{n,m});(\MATCHSym_{X^n} \Tensor \Id{X^n})
  \end{equation*}
  The mapping on 2-cells is again trivial. 
  Pseudofunctoriality  and square commutativity 
  follow as in the previous case.

  The bottom face is equal to the top one, so we now switch to 
  the side faces. Consider a graph morphism $g:G \to G'$.
  Commutativity of 
  \begin{equation*}
    \begin{tikzpicture}
      \node (G') at (-2,0) {$G'$}; 
      \node (G) at (-2,2) {$G$};
      \node (FG') at (0,0) {$\Free{G'}$}; 
      \node (FG) at (0,2) {$\Free{G}$};
      \node (Count') at (2,0) {$\Count$};
      \node (Count) at (2,2) {$\Count$};
      \draw[->] (G) -- node[left] {$g$} (G');
      \draw[->] (G) -- (FG);
      \draw[->] (G') -- (FG');
      \draw[->] (FG) -- node[left] {$\Free{g}$} (FG');
      \draw[->] (FG) -- (Count);
      \draw[->] (FG') -- (Count');
      \draw[transform canvas={xshift=-1pt}, -] (Count) -- (Count');
      \draw[transform canvas={xshift=1pt}, -] (Count) -- (Count');
    \end{tikzpicture}
  \end{equation*}
  Is just Lemma~\ref{lem: functor to count}. As for the square:
  \begin{equation*}
    \begin{tikzpicture}
      \node (Count) at (0,2) {$\Count$};
      \node (Count') at (0,0) {$\Count$};
      \node (Bzkp) at (2,2) {$\Bzkp{f(n,m)}$}; 
      \node (Bzkp') at (2,0) {$\Bzkp{f(n',m')}$}; 
      \draw[->] (Count) -- (Bzkp);
      \draw[->] (Count') -- (Bzkp');
      \draw[->] (Bzkp) -- (Bzkp');
      \draw[transform canvas={xshift=-1pt}, -] (Count) -- (Count');
      \draw[transform canvas={xshift=1pt}, -] (Count) -- (Count');
    \end{tikzpicture}
  \end{equation*}
  It is sufficient to define the pseudofunctor 
  $\Bzkp{f(n,m)} \to \Bzkp{f(n', m')}$ as sending 
  $X^n$ to $X^{n'}$, and the generating morphism
  \begin{equation*}
    (\Id{X^n} \Tensor \COPYSym_{X^{f(n,m)}} \Tensor \COPYSym_{X^m});
    (\Id{X^n \Tensor X^{f(n,m)}} \Tensor \sigma_{X^{f(n,m)},X^m} \Tensor \Id{X^m});
    (\Id{X^n} \Tensor S_{n,m} \Tensor T_{n,m});(\MATCHSym_{X^n} \Tensor \Id{X^n})
  \end{equation*}
  To:
  \small{
  \begin{equation*}
    (\Id{X^{n'}} \Tensor \COPYSym_{X^{f(n',m')}} \Tensor \COPYSym_{X^{m'}});
    (\Id{X^{n'} \Tensor X^{f(n',m')}} \Tensor \sigma_{X^{f(n',m')},X^{m'}} \Tensor \Id{X^{m'}});
    (\Id{X^{n'}} \Tensor S_{n',m'} \Tensor T_{n',m'});(\MATCHSym_{X^{n'}} \Tensor \Id{X^{n'}})
  \end{equation*}}
  According to this definition extensionally equal circuits 
  are sent to extensionally equal circuits, so 2-cells can be 
  defined in the obvious way. Commutativity of the square 
  is again true by definition. A slight modification of the 
  proof above allows us to also prove the commutativity
  of the following square:
  \begin{equation*}
    \begin{tikzpicture}
      \node (Count) at (0,2) {$\Count$};
      \node (Count') at (0,0) {$\Count$};
      \node (Bzkp) at (2,2) {$\Bgraph{G}$}; 
      \node (Bzkp') at (2,0) {$\Bgraph{G'}$}; 
      \draw[->] (Count) -- (Bzkp);
      \draw[->] (Count') -- (Bzkp');
      \draw[->] (Bzkp) -- (Bzkp');
      \draw[transform canvas={xshift=-1pt}, -] (Count) -- (Count');
      \draw[transform canvas={xshift=1pt}, -] (Count) -- (Count');
    \end{tikzpicture}
  \end{equation*}
  Now we focus on:
  \begin{equation*}
    \begin{tikzpicture}
      \node (G') at (-2,0) {$G'$}; 
      \node (G) at (-2,2) {$G$};
      \node (FG') at (0,0) {$\Free{G'}$}; 
      \node (FG) at (0,2) {$\Free{G}$};
      \node (Bpath') at (2,0) {$\Bpath{G'}$};
      \node (Bpath) at (2,2) {$\Bpath{G}$};
      \draw[->] (G) -- node[left] {$g$} (G');
      \draw[->] (G) -- (FG);
      \draw[->] (G') -- (FG');
      \draw[->] (FG) -- node[left] {$\Free{g}$} (FG');
      \draw[->] (FG) -- (Bpath);
      \draw[->] (FG') -- (Bpath');
      \draw[->] (Bpath) -- (Bpath');
    \end{tikzpicture}
  \end{equation*}
  Whose commutativity is obvious by defining the 
  pseudofunctor $\Bpath{G} \to \Bpath{G'}$ by sending 
  $X^n$ to $X^{n'}$, and every morpsism
  \begin{equation*}
    (\Id{X^n} \Tensor e);(\Id{X^n} \Tensor \COPYSym_{X^m});
    (\Id{X^n} \Tensor S_G \Tensor T_G);(\MATCHSym_{X^n} \Tensor \Id{X^n})
  \end{equation*}
  To:
  \begin{equation*}
    (\Id{X^{n'}} \Tensor \Free{g}(e));(\Id{X^{n'}} \Tensor \COPYSym_{X^{m'}});
    (\Id{X^{n'}} \Tensor S_G \Tensor T_G);(\MATCHSym_{X^{n'}} \Tensor \Id{X^{n'}})
  \end{equation*}
  2-cells mapping is again obvious.

  Finally, we focus on the squares:
  \begin{equation*}
    \begin{tikzpicture}
      \node (Bpath) at (0,2) {$\Bpath{G}$};
      \node (Bpath') at (0,0) {$\Bpath{G'}$};
      \node (Bgraph) at (2,2) {$\Bgraph{G}$}; 
      \node (Bgraph') at (2,0) {$\Bgraph{G'}$}; 
      \draw[->] (Bpath) -- (Bgraph);
      \draw[->] (Bpath') -- (Bgraph');
      \draw[->] (Bgraph) -- (Bgraph');
      \draw[->] (Bpath) -- (Bpath');
    \end{tikzpicture}
    \qquad
    \begin{tikzpicture}
      \node (Bgraph) at (0,2) {$\Bgraph{G}$};
      \node (Bgraph') at (0,0) {$\Bgraph{G'}$};
      \node (Bzkp) at (2,2) {$\Bzkp{f(m,n)}$}; 
      \node (Bzkp') at (2,0) {$\Bzkp{f(m',n')}$}; 
      \draw[->] (Bgraph) -- (Bzkp);
      \draw[->] (Bgraph') -- (Bzkp');
      \draw[->] (Bzkp) -- (Bzkp');
      \draw[->] (Bgraph) -- (Bgraph');
    \end{tikzpicture}
  \end{equation*}
  Whose commutativity is proven similarly. 
  All pseudofunctors involved in these squares 
  have already been defined previously. Commutativity 
  is obvious by tracking where generating morphisms get 
  mapped.
\end{proof} 
\end{document}